\documentclass[aps,pra,amsmath,amssymb,amsfonts,a4paper,10pt,twocolumn, 
superscriptaddress,nofootinbib]{revtex4-2}  

\usepackage[utf8]{inputenc}
\usepackage{rotating}
\usepackage{grffile}

\usepackage[xcolor]{changebar}  

\usepackage{url}

\usepackage[babel,autostyle]{csquotes}   
\usepackage[english]{babel}

\usepackage{framed}
\usepackage{lipsum}
\usepackage{floatpag}

\usepackage{graphicx}
\usepackage{siunitx}
\usepackage{mathtools}
\usepackage{dsfont}
\newcommand{\nicefrac}[2]{%
  \raise.5ex\hbox{$#1$}%
  \kern-.1em/\kern-.15em%
  \lower.25ex\hbox{$#2$}}

\usepackage{braket}
\usepackage{dcolumn} 
\usepackage{MnSymbol}
\usepackage{amsthm} 
\newcommand{\hH}{\hat{H}}
\newcommand{\tH}{\tilde{H}}
\newcommand{\bhH}{\hat{\mathbf{H}}} 
\newcommand{\hU}{\hat{U}}
\newcommand{\hV}{\hat{V}}
\newcommand{\hW}{\hat{W}}
\newcommand{\hA}{\hat{A}}
\newcommand{\hB}{\hat{B}}
\newcommand{\hbB}{\hat{\mathbf{B}}}
\newcommand{\hC}{\hat{C}}
\newcommand{\hD}{\hat{D}}
\newcommand{\hcalW}{\hat{\mathcal{W}}}
\newcommand{\tU}{\tilde{U}}
\newcommand{\UI}{\mathrm{UI}}
\newcommand{\Tr}{\mathrm{Tr}}
\newcommand{\St}{\mathrm{St}}
\newcommand{\bE}{\hat{E}}
\newcommand{\be}{\mathbf{e}}
\newcommand{\hE}{\hat{E}}
\newcommand{\expi}[1]{\exp\left(-i #1 \right)}

\newcommand{\diag}[1]{\mathop{}\!\mathrm{diag}\left( #1 \right)}

\newcommand{\balpha}{\boldsymbol{\alpha}}
\newcommand{\bc}{\mathbf{c}}
\newcommand{\bi}{\mathbf{i}}
\newcommand{\bj}{\mathbf{j}}

\newcommand{\bs}{\mathbf{s}}
\newcommand{\bN}{\mathbf{N}}
\newcommand{\bB}{\mathbf{B}}

\newcommand{\bigO}{\mathcal{O}}

\newcommand{\ft}[1]{{\frac{#1}{2}}}
\newcommand{\comb}[2]{{[\hH_#1, \hH_#2]}}
\newcommand{\comt}[3]{{[[\hH_{#1}, \hH_{#2}], \hH_{#3}]}}
\newcommand{\norm}[1]{{\lVert#1\rVert}}
\newcommand{\td}{\mathrm{d}}
\newcommand{\std}{\mathrm{std}}
\newcommand{\eig}{\mathrm{eig}}
\newcommand{\with}{\quad \mathrm{with} \quad}

\newcommand{\mini}{\mathrm{min}}
\newcommand{\maxi}{\mathrm{max}}
\newcommand{\ddfrac}[2]{\frac{\mathrm{d}{#1}}{\mathrm{d}{#2}}}

\definecolor{mylightblue}{rgb}{0.975,0.975,1}
\definecolor{myborders}{rgb}{0.3961,0.4157,0.4549}
\definecolor{mytitleinterior}{rgb}{0.898,0.9176,0.9569}
\definecolor{mygreen}{rgb}{0,0.6,0}
\definecolor{mygray2}{rgb}{0.5,0.5,0.5}
\definecolor{mymauve}{rgb}{0.58,0,0.82}

\usepackage{url}
\usepackage{ifthen}
\usepackage{caption}
\usepackage{subcaption}
\usepackage{tabularx}

\usepackage{listings}
\usepackage{hyperref}
\usepackage{cleveref}

\newcommand{\includepdf}[2]{%
  \edef\x{#1/#2}
  \includegraphics{\x.pdf}
}

\begin{document}
\title{Exponentiation of Parametric Hamiltonians via Unitary Interpolation}

\author{Michael Schilling}
\affiliation{Peter Grünberg Institute - Quantum Control (PGI-8), Forschungszentrum Jülich GmbH, Wilhelm-Johnen-Straße, 52425 Jülich, Germany}
\affiliation{Institute for Theoretical Physics, University of Cologne, Zülpicher Straße 77, 50937 Cologne, Germany}
\author{Francesco Preti}
\affiliation{Peter Grünberg Institute - Quantum Control (PGI-8), Forschungszentrum Jülich GmbH, Wilhelm-Johnen-Straße, 52425 Jülich, Germany}
\affiliation{Institute for Theoretical Physics, University of Cologne, Zülpicher Straße 77, 50937 Cologne, Germany}
\author{Matthias M. Müller}
\affiliation{Peter Grünberg Institute - Quantum Control (PGI-8), Forschungszentrum Jülich GmbH, Wilhelm-Johnen-Straße, 52425 Jülich, Germany}
\author{Tommaso~Calarco}
\affiliation{Peter Grünberg Institute - Quantum Control (PGI-8), Forschungszentrum Jülich GmbH, Wilhelm-Johnen-Straße, 52425 Jülich, Germany}
\affiliation{Institute for Theoretical Physics, University of Cologne, Zülpicher Straße 77, 50937 Cologne, Germany}
\affiliation{Dipartimento di Fisica e Astronomia, Università di Bologna, 40127 Bologna, Italy}
\author{Felix Motzoi}
\affiliation{Peter Grünberg Institute - Quantum Control (PGI-8), Forschungszentrum Jülich GmbH, Wilhelm-Johnen-Straße, 52425 Jülich, Germany}

\date{\today}

\begin{abstract}
    The effort to generate matrix exponentials and associated differentials, required to determine the time evolution of quantum systems, frequently constrains the evaluation of problems in quantum control theory, variational circuit compilation, or Monte-Carlo sampling.
    We introduce two ideas for the time-efficient approximation of matrix exponentials of linear multi-parametric Hamiltonians.
    We modify the Suzuki-Trotter product formula from an approximation to an interpolation schemes to improve both accuracy and computational time.
    This allows us to achieve high fidelities within a single interpolation step, which can be computed directly from cached matrices.
    We furthermore define the interpolation on a grid of system parameters, and show that the infidelity of the interpolation converges with $4^\mathrm{th}$
    order in the number of interpolation bins.
\end{abstract}

\maketitle

\section{Introduction}
The repeated construction of unitaries of linear parametric Hamiltonians is a cornerstone task in various research domains of quantum mechanics.
The ability to accurately and efficiently construct these unitaries enables us to predict and control the dynamics of quantum systems,
which is vital for both theoretical insights and practical applications ranging from
quantum simulation \cite{Lloyd1996, Georgescu2014} to quantum optimal control \cite{Koch2022} and quantum circuit compilation \cite{Martinez2016, Khatri2019, preti2022}.
Due the curse of dimensionality caused by the exponential growth of finite-dimensional representations of Hilbert spaces with problem size, even the generation of few unitaries can be computationally taxing.
Yet the afformentioned applications require the construction of many such unitaries.

A common challenge is calculating the time evolution operator of time-dependent Hamiltonians.
To address this, the typical strategy involves approximating these Hamiltonians using products of unitaries of sufficiently small time-steps.
Standard approaches to numerically compute these time-steps include the Trotter-product formula \cite{Trotter1959}, Dyson series \cite{Dyson1949},
Magnus expansion \cite{Magnus1954, Dalgaard2022}, Fer expansion \cite{Fer1958} and Runge-Kutta methods \cite{Runge1895, Kutta1901}.
In the case of quantum state propagation, one often turns to Krylov subspace methods \cite{Hochbruck1997}.

We consider problems with Hamiltonians of a few linear parameters,
that require the repeated construction of matrix exponentials (or their action onto states).
Such problems commonly occur within the context of quantum optimal control theory \cite{Koch2022},
where we shape system parameters in time, also referred to as pulse shaping,
to achieve the high fidelity operations required in high-precision quantum technologies. Methods such as
GRadient Ascent Pulse Engineering (GRAPE) \cite{Khaneja2005, motzoi2011, Fouquieres2011}, Krotov \cite{Morzhin2019},
DYNAMic Optimization (DYNAMO) \cite{dynamo1}, Chopped RAndom-Basis (CRAB) \cite{Caneva2011, Rach2015,CRABReview}, and
Single Optimization with multiple application (SOMA) \cite{preti2022} for generalized system parameter-dependent pulses,
require the repeated construction of matrix exponentials in order to propagate and optimize the propagation of the quantum systems. Another common application of multi-parameter Hamiltonians is in stochastic evolution \cite{Onorati2017} and other Monte Carlo methods \cite{Ceperley1986, Austin2012}.

Specifically, we introduce a matrix-product approach, which we call Unitary Interpolation (UI),
to interpolate between unitaries and approximate matrix exponentials to arbitrary precision.
We compare this new approach to existing schemes, namely exact matrix exponentiation via hermitian eigenvalue decomposition,
Trotterisation and in the case of state propagation Krylov exponentiation into a vector.
Standard approaches for the computation of the time-evolution of unitaries (vectors) have a complexity of $\mathcal{O}(d^3)$ \footnote{This can be reduced to $\mathcal{O}(d^{2.8})$ for matrix multiplication using the Strassen algorithm, which is not often used in practice as its advantages are limited to large matrices $d>1000$.}($\mathcal{O}(d^2)$),
where $d$ is the dimension of the systems Hilbert space, and with a large prefactor, e.g.~compared to matrix-matrix (matrix-vector) multiplication, especially when parallelelization is taken into account.
As such, they typically constitute a large bottleneck in classical simulation of quantum system.
We show that our approach can reduce the cost of repeated matrix exponentiation, compared to Trotter-based approaches and eigenvalue decomposition. The method allows for the inclusion of multiple system parameters, at the cost of requiring more precomputations. We also consider state evolution, and find further speed-ups compared with the Krylov subspace method. The interpolation of multiple unitaries can also be implemented directly on a quantum computer to perform Hamiltonian simulation \cite{Lloyd1996}, e.g. to implement a the unitary of a linear parametric Hamiltonian on a variational quantum circuit. \\

The paper is structured as follows: in Section \ref{sec:linear-parametric-hamiltonians} we define the unitary evolution of a time-dependent Hamiltonian with continuous linear parameters.
In Section \ref{sec:interpolation_between_two_unitaries} the interpolation between two arbitrary unitaries via displacement operations is introduced.
This method is applied to problems of one time-dependent parameter in Section \ref{sec:hamiltonians_with_a_single_control_parameter}.
We then generalize the displacement and the corresponding interpolation to problems with multiple time-dependent parameters in Section \ref{sec:hamiltonians_with_multiple_control_parameters}.
Furthermore, in Section \ref{sec:state-evolution}, we discuss the special case of state evolution using UI and in Section \ref{sec:caching} deduce best practices for caching. \\

\section{Unitaries of Linear Parametric Hamiltonians}\label{sec:linear-parametric-hamiltonians}
We consider the problem of computing unitaries $\hU(\bc) \in \text{U}(d)$ generated by linear parametric $d$-dimensional Hamiltonians $\hH(\bc)$
\begin{equation}
    \hU(\bc) = \exp(-i\hH(\bc)) \quad \text{with} \quad \hH(\bc) = \hH_0 + \bc^T \bhH, \label{eq:linear_parametric_hamiltonian}
\end{equation}
where $\bc = (c_1, \cdots, c_{n})^T$ is a vector of $n$ real amplitude-bound parameters $c_p \in [c_{p,\mini}, c_{p,\maxi}]$,
$\hH_0$ a $d$-dimensional drift Hamiltonian and $\bhH = (\hH_1, \cdots, \hH_{n})^T$ a $n$-dimensional vector of $d$-dimensional parameter Hamiltonians.

Let us first consider two types of problems, in which these kinds of Hamiltonians are commonly encountered.

\subsubsection{Time Evolution via Time-Slicing}
The time evolution $\hU(0, T)$ of quantum systems with time-dependent Hamiltonians
\begin{align}
    \tH(t) = \tH_0 + \sum_{p=1}^{\tilde{n}} \tilde{c}_p(t) \tH_p,
\end{align}
is represented by a product of smaller time-steps
\begin{align}
    \hU(0, T) = \prod_{j=0}^{m-1} \hU(t_j, t_{j+1}),
\end{align}
with $t_j = j \Delta t$ and $\Delta t = \frac{T}{m}$.
Each of the time-steps $[t_j, t_{j+1})$ is approximated by a unitary $\hU(t_j, t_{j+1}) \approx \tU(\bc_j)$ generated from a time-independent linear parametric Hamiltonian $\hH(\bc_j)$ of the form Eq.~\eqref{eq:linear_parametric_hamiltonian}.
In particular, the time-independent "surrogate" Hamiltonian $\hH(\bc_j)$ can be constructed from a truncated Magnus expansion \cite{Magnus1954, Dalgaard2022}.
The first $\tilde{n}$ terms of the expansion are given by the average parameter values in the interval
\begin{align}
    c_p = \frac{1}{\Delta t} \int_{t_j}^{t_{j+1}}\tilde{c}_p(t') dt'
\end{align}
and the normalized Hamiltonian elements $\hH_p = \frac{\tH_p}{\Delta t}$.
Higher order terms are given by the nested commutators of the original Hamiltonian elements $\hH_p$ and higher order integrals of the parameters $\tilde{c}_p(t)$, an instructive introduction can be found in \cite{Brinkmann2016}.

\subsubsection{Computation via Variational Quantum Circuits}
The classical simulation of variational quantum circuits requires the sequential multiplication of a given number $l$ of unitaries
\begin{align}
    \hU(\bc^{(1)}, ..., \bc^{(l)}) = \prod_{i=0}^{l-1} \hU^{(i)}(\boldsymbol{c}^{(i)})
\end{align}
where each $\hU^{(i)}$ is a quantum gate with parameters $\boldsymbol{c}^{(i)}$ acting on one or more qubits. When dealing with $n$-qubit collective gates, the computation of their unitary matrix, which needs to be evaluated at multiple parameter values -- e.g., in the context of variational quantum algorithms \cite{Cerezo2021} -- is usually computationally expensive, unless an analytical decomposition is available \cite{preti2022}. As parametric quantum gates can be represented in the form Eq.~\eqref{eq:linear_parametric_hamiltonian}, they can benefit from improved techniques to efficiently compute such unitaries $\hU^{(i)}(\bc^{(i)})$,
and the (analytical) gradients with respect to the parameters $\ddfrac{\hU^{(i)}(\bc^{(i)})}{c_p}$.

To speed-up such computations, we develop a technique to interpolate between unitaries,
and to create a cache of matrices from which we can construct such interpolations efficiently.
We first discuss the interpolation between two unitaries \ref{sec:interpolation_between_two_unitaries},
then apply the method to problems with only one parameter\ref{sec:hamiltonians_with_a_single_control_parameter},
generalize to multiple system parameters \ref{sec:hamiltonians_with_multiple_control_parameters}
and finally discuss the special case of state evolution \ref{sec:state-evolution}. We assess the accuracy of the unitaries via the average gate infidelity $I_\mathrm{avg}=\frac{d}{d+1} -\frac{\vert\mathrm{Tr}(\hU_\mathrm{approx} \hU_\mathrm{exact}^\dagger)\vert^2}{d(d+1)}$ between the exact $\hU_\mathrm{exact}$ and approximated unitaries $\hU_\mathrm{approx}$, see Appendix~\ref{sec:average_infidelity}. We perform analogous assessments for quantum states.

\section{Interpolation between two Unitaries} \label{sec:interpolation_between_two_unitaries}
\begin{figure}[t!]
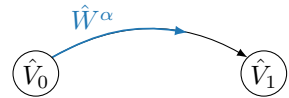

    \centering
    \includepdf{Grid}{displacement_W_1D}
    \caption{Interpolation between unitaries $\hV_0$ and $\hV_1$ via the displacement operation $\hW^\alpha$ with $\alpha \in [0,1]$ (here $\alpha=\nicefrac{3}{4}$),
        which transforms $\hV_0$ into $\hV_1$ for $\alpha=1$. \label{fig:displacement_W_1D}}
\end{figure}
Any two unitary operators $\hV_0, \hV_1 \in \mathds{C}^{d\times d}$ can be written in terms of generators $\hA$ and $\hB$ as $\hV_0 = \exp(-i\hA)$ and $\hV_1 = \exp(-i(\hA + \hB))$,
where $\hB$ is the difference generator.
We wish to efficiently construct operators of the form $\exp(A+\alpha B)$, where $A$ and $B$ are anti-hermitian operators and $\alpha \in [0,1]$.

To this end we can construct an interpolation between the unitaries, from the following product formula of the unitaries
\begin{equation}
    \hV(\alpha) = \underbrace{\left(\hV_1 \hV_0^\dagger \right)^\alpha}_{=\hW^\alpha} \hV_0. \label{eq:geodesic}
\end{equation}
We call $\hW = \left(\hV_1 \hV_0^\dagger \right)$ the displacement operator and $\alpha \in (0,1)$ the interpolation parameter, see Fig.~\ref{fig:displacement_W_1D}.
We find $\hV(\alpha)$ to be an \textit{interpolation}, as it reproduces the original operators on the edges of the open interpolation interval:
\begin{equation}
    \hV(0) = \hV_0 \with \hV(1) = \hV_1. \label{eq:1d_interpolation_exact_at_borders}
\end{equation}
The interpolation $\hV(\alpha)$ is furthermore a \textit{unitary} operator, as any product of unitaries is itself unitary,
hence the name Unitary Interpolation (UI).

The first order BCH expansion is constructed from the sum of the generators (using a sign change for the hermitian conjugate), revealing
\begin{equation}
    \hV(\alpha) = \exp(-i(\hA + \alpha \hB) + \cdots). \label{eq:ui_bch_first_order}
\end{equation}
To first order, this corresponds to the desired operator $\hV(\alpha) = \exp(-i(\hA + \alpha \hB))$.
Deviations from the desired form are given by higher order terms in the BCH expansion.
We can investigate these terms by studying the properties of the displacement operator $\hW$.

The BCH expansion of the displacement operator $\hW$ exponentiated with the interpolation parameter $\alpha$ is given by
\begin{equation}
    \begin{aligned}
        \hW = (\hV_1 \hV_0^\dagger)^\alpha & = \left(\exp(-i(\hA + \hB)) \exp(i\hA)\right)^\alpha                 \\
                                           & = \exp\left(\alpha(-i\hB - \frac{1}{2}[ \hA, \hB] + \cdots )\right)  \\
                                           & = \exp\left(\alpha \sum_i \hC_i \right), \label{eq:displacement_bch}
    \end{aligned}
\end{equation}
where we generically denoted the nested commutators of $\hA$ and $\hB$ as $\hC_i$, e.g. $\hC_1 = -i\hB$, $\hC_2=\frac{1}{2}[ \hA, \hB]$ and $\hC_3=\frac{i}{12}[\hB, [\hA, \hB]]$.
The nested commutators $\hC_i$ are at least linear in $\hB$, as commutators built from only one operator ($\hA$) would vanish.
Furthermore every term is linear in $\alpha$. Hence, for every term the order of $\hB$ is at least as large as the order of $\alpha$.

For the interpolation -- see Eq.~\eqref{eq:geodesic} -- we perform a similar generic BCH expansion for the undesired higher order terms
\begin{equation}
    \hV(\alpha) = \exp\Big(\overbrace{-i(\hA + \alpha \hB)}^{\mathrm{Desired Terms}} + \overbrace{\sum_i p_i(\alpha)\hD_i}^{\mathrm{Undersired Terms}}\Big). \label{eq:ui_bch}
\end{equation}
The undesired terms are represented by a sum of polynomials $p_i(\alpha)$ and corresponding nested commutators $\hD_i$ of the generators $\hA$ and $\hB$,
e.g. $\hD_1 = [\hB, [\hA, \hB]]$. The operators $\hD_i$ are can similarly be constructed as nested commutators of $\hA$ and the $\hC_i$ from Eq.~\eqref{eq:displacement_bch}.
For example, $\hD_1$ is constructed both via $[\hC_1, [\hA, \hC_1]]$ and directly from $\hC_3$. The path to $\hD_1$ via $\hC_3$ contains only one $\hC$ term
and therefore contributes only a linear $\alpha$ to the polynomial $p_1(\alpha)$, wheras the path via $[\hC_1, [\hA, \hC_1]]$ contains two $\hC$ terms
and contributes a quadratic term in $\alpha$ to $p_1(\alpha)$.

Each of the undesired terms must vanish for $\alpha=0$ and $\alpha=1$, hence the polynomials $p_i(\alpha)$
must be at least quadratic in $\alpha$ (so as to have the two roots), meaning that one of the constructions of $\hD_i$ must contain at least two $\hC$ terms.
Therefore every term $\hD_i$ is at least quadratic in $\hB$. The infidelity is quadratic in the undesired terms,
it scales with the $4^\mathrm{th}$ power of the difference operator $\hB$.

\subsection{Construction from Cached Matrices} \label{sec:geodesic_from_cache}
In order to construct the interpolations given in Eq.~\eqref{eq:geodesic} repeatedly and quickly,
we decompose the displacement operator $\hW$ via Shur decomposition
\begin{equation}
    \hW = (\hV_1 \hV_0^\dagger) = \hat{L}\exp(-i\bE)\hat{L}^\dagger
\end{equation}
into the eigenvector matrix $\hat{L}$ and the diagonal eigenvalue matrix $\exp(-i\bE)$,
which we cache as a vector $\mathbf{E}=\mathrm{diag}(\bE)$.
For the right side we cache the product $\hat{R} = \hat{L}^\dagger \hV_0$,
so that we can rewrite the interpolation as the product
\begin{equation}
    \hV(\alpha) = \hat{L} \exp(-i\be \alpha) \hat{R}. \label{eq:geodesic_from_cache}
\end{equation}
The matrix exponential in the middle can be performed as a row-wise scaling operation on the right side matrix $\hat{R}$,
so that the relevant complexity reduces to a single matrix multiplication.

\begin{figure}[t!]
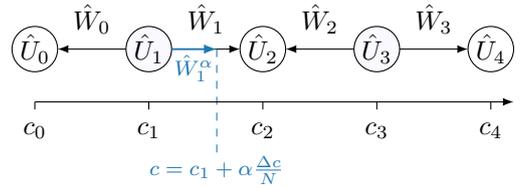

    \centering
    \includepdf{Grid}{grid_1d_displacement}
    \caption{The interpolation of a unitary using a single system parameter $c$ using multiple interpolation bins.
        Interpolation is performed from the odd-indexed unitaries ($\hU_1$, $\hU_3$) towards the adjacent even indexed unitaries ($\hU_0$, $\hU_2$, $\hU_4$),
        an odd-centered interpolation grid. \label{fig:grid_1D}}
\end{figure}

\section{Unitary Interpolation of a single System Parameter} \label{sec:hamiltonians_with_a_single_control_parameter} 
We have seen, that the UI allows us to approximate unitaries of the form
$\exp(-i(\hA + \alpha \hB))$, constructed from Hermitian operators $\hA$ and $\hB$.
Importantly we found that the corrections scale with the $4^\mathrm{th}$
power of the difference operator $\hB$, which we will use to reduce the infidelity of the interpolation.
We will now apply the approach to a Hamiltonian of a single parameter and reduce the size of $\hB$
to reduce the contributions of the undesired terms.

Let us consider the unitary $\hU(c)$ generated by a Hamiltonian $H(c)$ of a single bounded parameter $c \in [c_\mini, c_\maxi]$
\begin{equation}
    \hU(c)= \exp(-i\hH(c)) \quad \text{with} \quad \hH(c)=\hH_0+c\hH_1. \label{eq:hamiltonian_1d}
\end{equation}
In order to repeatedly calculate unitaries $\hU(c)$ for different values of $c$, we separate the parameter interval into a suitably\footnote{We discuss optimal binning in \ref{sec:optimal_caching}}
chosen number $N$ of (equidistant) interpolation intervals/bins, see Fig.~\ref{fig:grid_1D}.
The bins are separated by the interpolation $N+1$ grid vertices $c_i$ for $i \in \{0, \cdots, N\}$,
\begin{equation}
    c_i=c_\mini + i\frac{\Delta c}{N}  \with \Delta c = c_\maxi-c_\mini,  \label{eq:interpolation_param_1d}
\end{equation}
for which we compute the corresponding grid unitaries
\begin{equation}
    \hU_i = \hU(c_i) = \expi{\hH(c_i) \Delta t}. \label{eq:grid_unitaries_1d}
\end{equation}
To perform the UI for a system parameter $c$, we decompose it into a grid vertex $c_i$
and an interpolation parameter $\alpha(c)$
\begin{equation}
    c = c_i + \alpha(c) \frac{\Delta c}{N},  \label{eq:control_param2interpolation_param_1d}
\end{equation}
where $\frac{\Delta c}{N}$ is the width of a single interpolation bin.
The interpolation is performed between adjacent grid unitaries $\hU_i$ and $\hU_{i\pm1}$,
starting from the closest odd-indexed unitary $\hU_i$ with $i \mod 2 = 1$ towards the closest even-indexed
unitary $\hU_j$ with $j = i + \mathrm{sign}(\alpha(c))$, where $\alpha(c) = ( c-c_i ) \frac{N}{\Delta c} \in (-1, 1)$.
The interpolation $\hU_\UI(c)$ is then constructed using a bin specific displacement operator
\begin{equation}
    \hU_\UI(c) = {\underbrace{\left(\hU_j \hU_i^\dagger \right)}_{=\hW_k}}^{|\alpha(c)|} \hU_i. \label{eq:ui_1d}
\end{equation}
This displacement operator $\hW_k$ is uniquely indexed by the
minimum of the indexes involved $k=\min(i,j)$. We take the absolute value of the interpolation parameter $|\alpha(c)|$,
because by definition it can be negative now, as the interpolation is performed from the odd-indexed unitaries towards the even-indexed unitaries.

\begin{figure}[t]
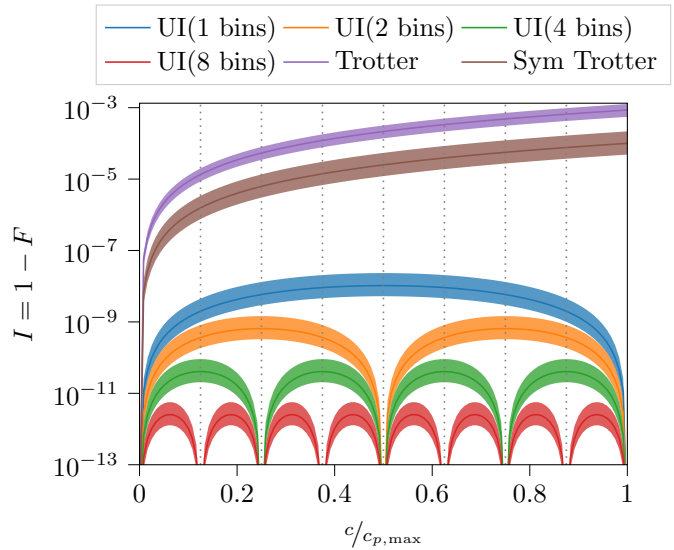

    \centering
    \includepdf{Infidelity}{one_d_figures_combined}
    \caption{The target-unitary infidelity of a single parameter problem using (symmetric) Trotterisation
        and Unitary Interpolation with 1, 2, 4 or 8 bins.
        The plot shows one standard deviation of the Infidelity sampled using random $d=16$ dimensional Hamiltonians
        with  $\std(\bE_p)=\nicefrac{\pi}{2}$ and the maximum parameter amplitude $c_{p, \mathrm{max}}=0.025$.
        A single UI interval improves the Infidelity and accuracy of the gradients of the symmetric Trotterisation by more than $3$ orders of magnitude.
        \label{fig:1d_infidelity}}
\end{figure}

\subsection{Accuracy Analysis} \label{sec:1d_error_analysis}
Using the separation of the the system parameter $c$ into grid vertex $c_i$ and interpolation parameter $\alpha(c)$, Eq.~\eqref{eq:interpolation_param_1d},
we can separate the target Hamiltonian in Eq.~\eqref{eq:hamiltonian_1d}
\begin{equation}
    \hH(c)\Delta t = \underbrace{\left(\hH_0 + c_i \hH_1\right)}_{=\hA_i} + \alpha(c) \underbrace{\frac{\Delta c}{N} \hH_1}_{=\hB}. \label{eq:hamiltonian_1d_interpolation}
\end{equation}
into the operators $\hA_i$ and $\hB$, corresponding to the (interval specific) generators of our analysis in \ref{sec:interpolation_between_two_unitaries}.
The binning allows us to rescale the difference operator $\hB$, to achieve our target infidelity.
We showed in \ref{sec:interpolation_between_two_unitaries}, that the infidelity scales with the $4^\mathrm{th}$ power of the difference operator $\hB$,
so that the infidelity of our binned interpolation is proportional to $\bigO\left(\frac{\Delta c^4}{N^4}\right)$.
As a consistency check we give a symbolically derived BCH expansion in Appendix \ref{sec:ui_1d_bch}.

By comparison, the infidelity of $N_T$ Suzuki-Trotter steps \cite{Trotter1959}, see Eq.~\eqref{eq:trotter_bch}, scales quadratically
$\bigO\Big(\frac{c^2}{N_T^2}\Big)$ both in the number of steps $N_T$ and the parameter amplitude $c$,
see the BCH expansion in \ref{sec:trotter_bch}. Similar to the $N$ bins of the UI,
the number of Trotter steps $N_T$ also rescales the difference operator $\hB= \frac{\Delta c}{N} \hH_1$.

We can improve on this by performing $N_S$ Symmetric Trotter steps (also known as split steps) \cite{Strang1968},
for which the infidelity scales as $\bigO\Big( \frac{c^2}{N_S^4}\Big)$, see Eq.~\eqref{eq:strang_bch}, quadratically in the parameter amplitude $c$,
but $4^\mathrm{th}$ order in the number of steps $N_S$.

As a side remark, the symmetrisation approach used in the Symmetric Trotter method, which removes all odd-ordered BCH correction terms of $N_S$,
does not translate to significant improvements for the UI. The symmetric (split-step) UI, derived in \ref{sec:sym_ui_bch}, still has $4^\mathrm{th}$ order correction terms in the infidelity
as the odd-ordered BCH correction terms are already zero, leading only to a constant factor reduction of the infidelity.
Attempts to increase the order of convergence with an interpolation should instead rely on increasing the number of interpolation roots (points at which the interpolation is exact), see the argument in Sec.~\ref{sec:interpolation_between_two_unitaries}

An overview of the leading order scaling of the different methods, both in the generators
and the infidelity is given in Tab.~\ref{tab:convergence_orders_by_method}.

In Fig.~\ref{fig:1d_infidelity} we show the infidelity as a function of the system parameter $c$ for a single parameter problem.
We compare $N \in [1,2,4,8]$ interpolation bins and a single Trotter and Symmetric Trotter step, as a single step requires same number of computational steps as the UI.
Every doubling in the number of bins leads to a reduction of the infidelity by a factor of $2^4=16$.

\begin{table}[thbp]
    \centering
    \caption{Convergence orders of the generators and the infidelity for different methods. The expansion was performed in orders of
        $\bigO\left( \frac{(\Delta c)^r}{N^s}\right)$, the table lists the exponents $r$ and $s$ for the leading order terms.
        In Trotterisation schemes time is split into $N$ steps, equivalently in UI, parameter amplitudes
        are split into $N$ intervals (bins). \label{tab:convergence_orders_by_method}}
    \begin{tabular}{l c c | c c}
        \phantom{Method} & \multicolumn{2}{c}{Generators \phantom{-}} & \multicolumn{2}{c}{Infidelity \phantom{-}}                                          \\ %
        \hline
        \phantom{Method} & $t$ for $(\Delta c)^t$                     & $s$ for $N^s$                              & $r$ for $(\Delta c)^r$ & $s$ for $N^s$ \\
        \hline
        Trotter          & 1                                          & 1                                          & 2                      & 2             \\
        Sym. Trotter     & 1                                          & 2                                          & 2                      & 4             \\
        UI               & 2                                          & 2                                          & 4                      & 4             \\
        Sym. UI          & 2                                          & 2                                          & 4                      & 4             \\
        \hline
    \end{tabular}
\end{table}

\subsubsection{Construction from Cached Matrices}
In analogy to the caching proposed in \ref{sec:geodesic_from_cache},
we can cache bin dependent matrices $\hat{L}_k$, $\hat{R}_k$ and $\bE_k$,
from the Shur decomposition of the displacement operators $\hW_k$
\begin{equation}
    \hW_{k} = (\hU_j \hU_i^\dagger) = \hat{L}_{k}\exp(-i\bE_{k})\hat{L}_{k}^\dagger, \label{eq:shur_decomposition}
\end{equation}
with $k=\min(i,j)$ is the minimum of the involved grid indices, $i$ being the odd index and $j=i\pm 1$ the even index.
We furthermore define $\hat{R}_k = \hat{L}^\dagger_k \hU_i$,
so that the interpolation can be calculated via
\begin{equation}
    \hU_\UI(c) = \hat{L}_k \expi{\bE_k |\alpha(c)| } \hat{R}_k. \label{eq:ui_1d_product}
\end{equation}
The product of the diagonal exponential $\expi{\bE_k |\alpha(c)| }$ with the right side matrix $\hat{R}_k$ can be performed
as a row-wise scaling operation, which has complexity $\mathcal{O}(d^2)$.
The relevant complexity then reduces to a single matrix multiplication.

\subsection{Gradients} \label{sec:1d_gradients}
Gradient based optimization techniques, eg. Gradient descent \cite{Fletcher2013}, noise propagation \cite{dalgaard2022b}, as well as exact \cite{Dalgaard2020} and approximate second order optimizers
such as Adam \cite{Kingma2014} and L-BFGS \cite{Nocedal1980} require the gradients of the objective function with respect to the system parameters.
Our interpolation scheme allows for the efficient calculation of the gradient of the unitaries with respect
to the system parameter
\begin{equation}
    \ddfrac{\hU_\UI(c)}{c} =  -\text{sign}(\alpha) \hat{L}_k \frac{i\bE_k N }{\Delta c} \expi{\bE_k |\alpha(c)| } \hat{R}_k. \label{eq:1d_gradients}
\end{equation}
This removes the need for either (semi-) automatic differentiation \cite{Goerz2022, Leung2017} or
analytical gradients from diagonalisation \cite{Fouquieres2011, Kuprov2009, Lewis2001}.

\section{Unitary Interpolation of Multiple Parameters}\label{sec:hamiltonians_with_multiple_control_parameters} 
We generalize the one-parameter UI to interpolate unitaries of linear Hamiltonians with $n$ system parameters -- see Eq.~\eqref{eq:linear_parametric_hamiltonian}.
First we generalize the one-parameter displacement operators to the $n$-parameter case in \ref{sec:nd_displacement_operators},
afterwards we construct a regular tiling of $n$-dimensional parameter spaces,
and generalize the odd-indexed grid to an odd-summed lattice \ref{sec:nd_voronoi_cells}.
Eventually we discuss the resulting $n$-parameter interpolation scheme in \ref{sec:nd_interpolation}.

\subsection{Displacement Operators in n-parameter Space} \label{sec:nd_displacement_operators}
Let us now consider $n+1$ unitaries in $U(d)$, a seed unitary $\hV_0 = \exp(-i \hA)$ and $n$ unitaries $U_i=\exp(-i (\hA + \hB_i))$,
with the difference operator $\hB_i$ being the difference between the generators of $\hV_0$ and $\hV_i$.
We construct $n$ one-parameter displacement operators $\hW^{(i)}$ to displace the seed unitary $\hV_0$ to the the unitaries $\hV_i$
\begin{equation}
    \left(\hW^{(i)}\right)^{\alpha_i} = \left(\hV_i \hV_0^\dagger \right)^{\alpha_i}. \label{eq:nd_displacement_operator}
\end{equation} 
The $n$-parameter displacement operator $\hcalW$ is constructed as a product of the one-parameter displacement operators
\begin{equation}
    \hcalW(\balpha) = \prod_{i=1}^n \left(\hW^{(i)}\right)^{\alpha_i}, \label{eq:nd_displacement_operator_product}
\end{equation}
with the interpolation parameter vector $\balpha = (\alpha_1, \dots, \alpha_n)^T$ of the interpolation parameters $\alpha_i$.
The action of this displacement operator $\hcalW(\balpha)$ on the seed unitary $\hV_0$ performs an $n$-parameter interpolation
\begin{equation}
    \hV(\balpha) = \hcalW(\balpha) \hV_0 = \left(\prod_{i=1}^n \left(\hV_i \hV_0^\dagger \right)^{\alpha_i}\right) \hV_0, \label{eq:nd_interpolation_general}
\end{equation}
which reproduces the unitaries used in its construction for
\begin{equation}
    \hV(\boldsymbol{0}) = \hV_0 \quad \mathrm{and} \quad \hV(\be_i) = \hV_i, \label{eq:nd_interpolation_exact_at_borders}
\end{equation}
where $\be_i$ is the unit vector in direction $i$.
The first order BCH expansion of the interpolation in Eq.~\eqref{eq:nd_interpolation_general} can again be calculated from the sum of the generators,
weighed by the interpolation parameters $\alpha_i$, so that we can write
\begin{equation}
    \hV(\balpha) = \exp\Big(\underbrace{-i(\hA + \balpha^T \bB)}_{\text{Desired Terms}} + \sum_j p_j(\balpha)\hD_j \Big). \label{eq:ui_bch_nd}
\end{equation}
The first order expansion reveals a linear equation of the interpolation parameters $\balpha$ and the difference operators $\bB = (\hB_1, \dots, \hB_n)^T$,
of a suitable form to approximate unitaries of $n$-parameter Hamiltonians Eq.~\eqref{eq:linear_parametric_hamiltonian}. These are the desired terms.
The higher order (undesired terms) are in analogy to Eq.~\eqref{eq:ui_bch} represented by a sum of polynomials $p_j(\balpha)$ and
commutators $\hD_j$ of the generators $\hA$ and $\hB_i$, i.e. $\hD_1 = [\hB_1, \hB_2]$.

The undesired terms must vanish for $\balpha=\mathbf{0}$ and $\balpha=\be_i$ for all $i$.
This (again) requires the polynomials $p_j(\balpha)$ to be at least quadratic in the $\alpha_i$ (i.e. $\alpha_i \alpha_j$),
to satisfy the $n+1$ roots.\footnote{Proof: Any linear polynomial can be written as $p(\balpha)=\sum_{i=1}^n c_i \alpha_i + c_0$.
    From the seed root $p(\mathbf{0}) = 0$ follows $c_i = 0$. From the $n$ roots $p(\hat{e}_i) = 0$ follows $c_i = 0$ for all $i$.
    So that only the trivial linear polynomial remains.}
In addition, using the fact that the every term in the displacement operators defined in Eq.~\eqref{eq:displacement_bch} is linear in
$\alpha_i$ and at least linear in $\hB_i$ it must follow
that the commutators $\hD_j$ must be at least quadratic in the $\hB_i$ (i.e. $[\hB_i, \hB_j]$), as was the case in the one-parameter UI.

\subsection{Tiling of the n-dimensional Parameter Space} \label{sec:nd_tiling}
In order to construct the UI of a linear $n$-parameter Hamiltonian in Eq.~\eqref{eq:linear_parametric_hamiltonian},
up to a target infidelity, for amplitude constrained system parameters $c_p \in [c_{p, \mathrm{min}}, \, c_{p, \mathrm{max}}]$,
we divide the $n$ dimensional parameter space $V$ into a regular tiling of interpolation cells $V_\bi$,
to uniquely associate every point $\bc \in V$ with an interpolation cell $V_\bi$.

We divide the system parameter (hyper-) volume into a simple cubic (hyper-) lattice of $\bN=(N_1,\dots,N_{n})$ bins\footnote{Chosen to achieve a target infidelity, we discuss an approach for binning in \ref{sec:optimal_caching}}
along the $n$ parameter directions, with the lattice-site parameter amplitudes
\begin{equation}
    \bc_\bi = (c_{1 i_1}, \dots, c_{n i_n})^T \quad \mathrm{with} \quad c_{p i_p}=c_{p,\mini} +i_p \frac{\Delta c_p}{N_p}, \label{eq:nd_lattice_site}
\end{equation}
identified via the lattice-site indexes $\bi=(i_1,\dots,i_n)$, where $i_p \in \{0,\dots,N_p\}$ and the lattice sizes $\Delta c_p = c_{p,\mathrm{max}} - c_{p,\mathrm{min}}$.
Correspondingly we define the lattice-site unitaries $\hU_\bi$
\begin{equation}
    \hU_{\bi} = \expi{\left(\hH_0+\bc_\bi^T \bhH \right)}, \label{eq:nd_lattice_unitaries}
\end{equation}
using the lattice-site parameter amplitudes $\bc_\bi$. The displacement operators are defined from adjacent lattice-site unitaries $\hU_{\bi}$ and $\hU_{\bi+\be_p}$.

\subsubsection{Voronoi Cells} \label{sec:nd_voronoi_cells}
The interpolation (hyper-) volumes are constructed as a regular Voronoi grid \cite{Voronoi1908}.
To construct the Voronoi grid, we choose the odd-summed lattice-sites $\bc_\bi$ with $\{\bi \mid (\sum_{k=1}^n i_k)\mod 2 = 1\}$ as seed points for the Voronoi cells $V_\bi$.
All points $\bc$ that are closer to a seed point $\bc_\bi$ than any other odd-summed grid points $\bc_{\bj \neq \bi}$ are part of the Voronoi cell $V_\bi$
\begin{equation}
    V_\bi = \left\{ \mathbf{x} \in \mathds{R}^d \mid \lVert \mathbf{x} - \mathbf{c}_\bi \rVert = \min_{\bj} \lVert \mathbf{x} - \mathbf{c}_\bj \rVert \right\} \label{eq:voronoi_cell}
\end{equation}
For points with two or more odd-summed lattice-sites at equal distance,
any choice between the neighbors can be made, the specific method we employ is discussed in App.~\ref{sec:voronoi_cell},
which discusses the precise approach we used to determine the Voronoi cell for system parameters $\bc$.

\begin{figure}
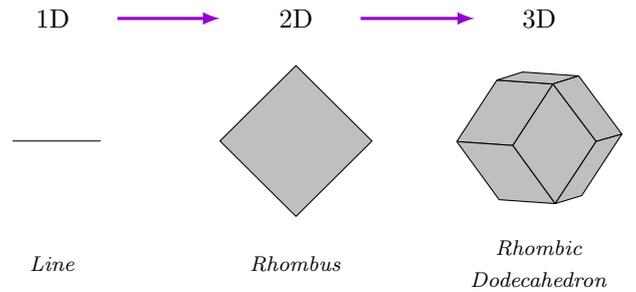

    \centering
    \includepdf{Grid}{polytopes}
    \caption{Examples of interpolation lines, areas (Rhombus) and volumes (Rhombic Dodecahedron) for 1, 2 and 3 parameter UI. \label{fig:interpolation_volumes}}
\end{figure}

In 1D we interpolate on line segments, in 2D on rhombi and in 3D on rhombic dodecahedra, see Fig.~\ref{fig:interpolation_volumes}.

\begin{figure}[t]
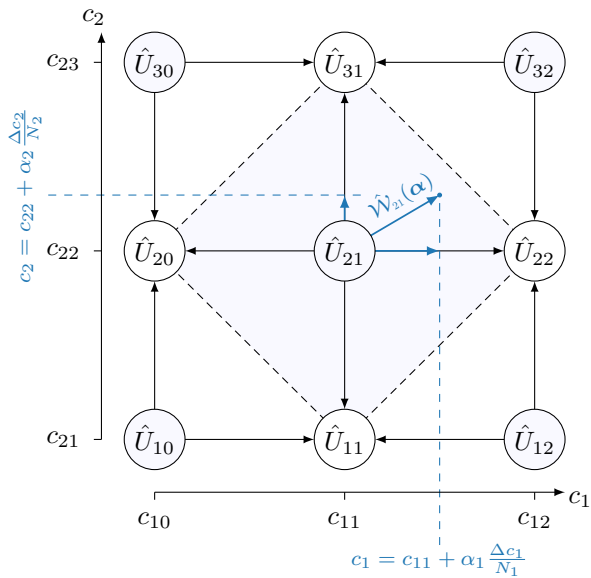

    \centering
    \includepdf{Grid}{new_decomposition_displacement}
    \caption{ A 2x2 Grid \textit{segment} extracted from a 2 parameter interpolation grid.
        The arrows point from the seed points along the approximate trajectory achieved by the one-parameter displacements.
        The index vectors are expanded into their two components, i.e. $\bi=(2,1) \rightarrow 21$.
        The interpolation is carried out from $\hU_{21}$ via the displacement operator $\hat{D}_{21}(\alpha)$. \label{fig:2d_binning}}
\end{figure}

\subsection{Unitary Interpolation}\label{sec:nd_interpolation}
In order to approximate a unitary $\hU(\bc)$ , we first decompose the system parameters $\bc$ into the odd-summed lattice-site $\bc_\bi$
(the seed of the corresponding Voronoi cell) and interpolation parameters $\balpha = (\alpha_1, \dots, \alpha_n)^T$, via
\begin{equation}
    \bc = \bc_\bi + \balpha \frac{\Delta \bc}{\bN}. \label{eq:control_param2interpolation_param_nd}
\end{equation}
The interpolation parameter term $\balpha \frac{\Delta \bc}{\bN}$ describes the location with respect to the seed point $\bc_\bi$.

The interpolation is achieved via the apropriate displacement operation $\hcalW_{\bi}(\balpha)$
\begin{equation}
    \hcalW_{\bi}(\balpha) = \prod_{p=1}^{n} \left( \hU_{\bi+bs_p\be_p} \hU_\bi^\dagger\right)^{|\alpha_p|},  \label{eq:nd_displacement_grid}
\end{equation}
where we use the signed index shift vector $\bs_p = \mathrm{sign}(\alpha_p) \be_p$ for more clarity.
The sign of the interpolation parameter $\alpha_p$ determines the direction in which we interpolate from the seed point $\bc_\bi$ (up or down the grid).
The product is used from the right to the left $\prod_{i=1}^s x_i = x_s \cdots x_2 x_1$.  The interpolation is then constructed from
\begin{equation}
    \hU_\UI (\bc) = \hcalW_{\bi}(\balpha) \hU_\bi= \left(\prod_{p=1}^{n} \left( \hU_{\bi+bs_p} \hU_\bi^\dagger\right)^{|\alpha_p|}\right) \hU_\bi,  \label{eq:nd_ui}
\end{equation}
The principle of an interpolation from a seed point, towards a point within its Voronoi cell is shown in Fig.~\ref{fig:2d_binning}, for a 2 parameter interpolation grid .

\begin{figure}[t]
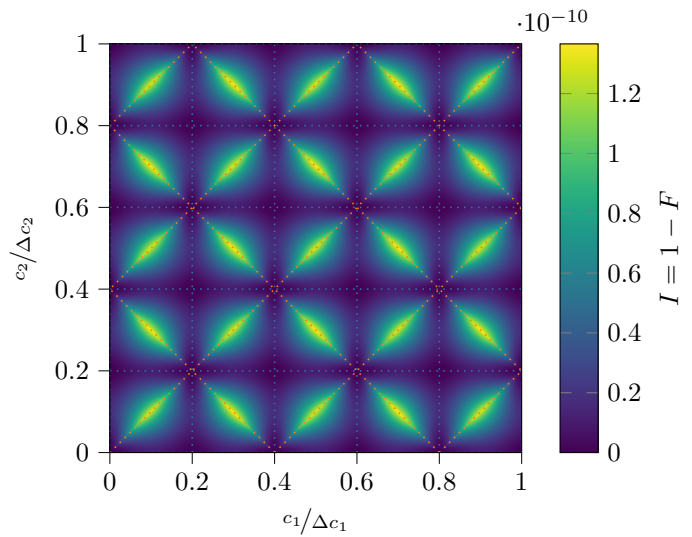

    \centering
    \includepdf{Infidelity}{two_d_5_bins_infidelity_comparison_UI}
    \caption{Infidelity map of A 2 parameter UI using $5 \times 5$ bins and a random Hamiltonian,
        see App.~\ref{sec:random_hamiltonians}, with $\std(\bE_p)=\nicefrac{\pi}{2} \, \forall p \in  \{1,\dots,n\}$ and $\Delta c_p=0.025$.
        The UI achieves an infidelity of $I(\bc) < 1.37\num{e-10}$.\label{fig:2d_infidelity} }
\end{figure}

\subsection{Accuracy and Performance}\label{sec:nd_error_analysis}
The decomposition given in Eq.~\eqref{eq:control_param2interpolation_param_nd} carries over the the Hamiltonian
\begin{equation}
    \hH(c) \Delta t = \underbrace{\hH_0 + \bc_\bi^T \bhH}_{=\hA_\bi} + \balpha^T \underbrace{\frac{\Delta \bc}{\bN} \bhH}_{=\hbB}, \label{eq:hamiltonian_nd_interpolation}
\end{equation}
and is conveniently separated into the seed Hamiltonian $A_\bi$ and the difference operator $\hbB$, which scales with the bin sizes $\frac{\Delta \bc}{\bN}$.
Our analysis in \ref{sec:nd_displacement_operators} proved, that in the same way as the one-parameter interpolation,
the infidelity scales with the $4^\mathrm{th}$ power of the difference operator $\hbB$. Together with Eq.~\eqref{eq:hamiltonian_nd_interpolation} we conclude,
that the infidelity is proportional to $\bigO\left(\frac{\Delta \bc^4}{\bN^4}\right)$ and that the results in Tab.~\ref{tab:convergence_orders_by_method} still hold valid.
The BCH expansion for $n$-parameter interpolations is given in \ref{sec:ui_nd_bch}.

The leading order correction is maximal at $\alpha_p = \nicefrac{1}{2} \, \forall p \in \{1,\dots,n\}$.
In Fig.~\ref{fig:2d_infidelity} we show the infidelity as a function of the system parameters for a 5x5 grid.

\begin{figure*}[hbtp]
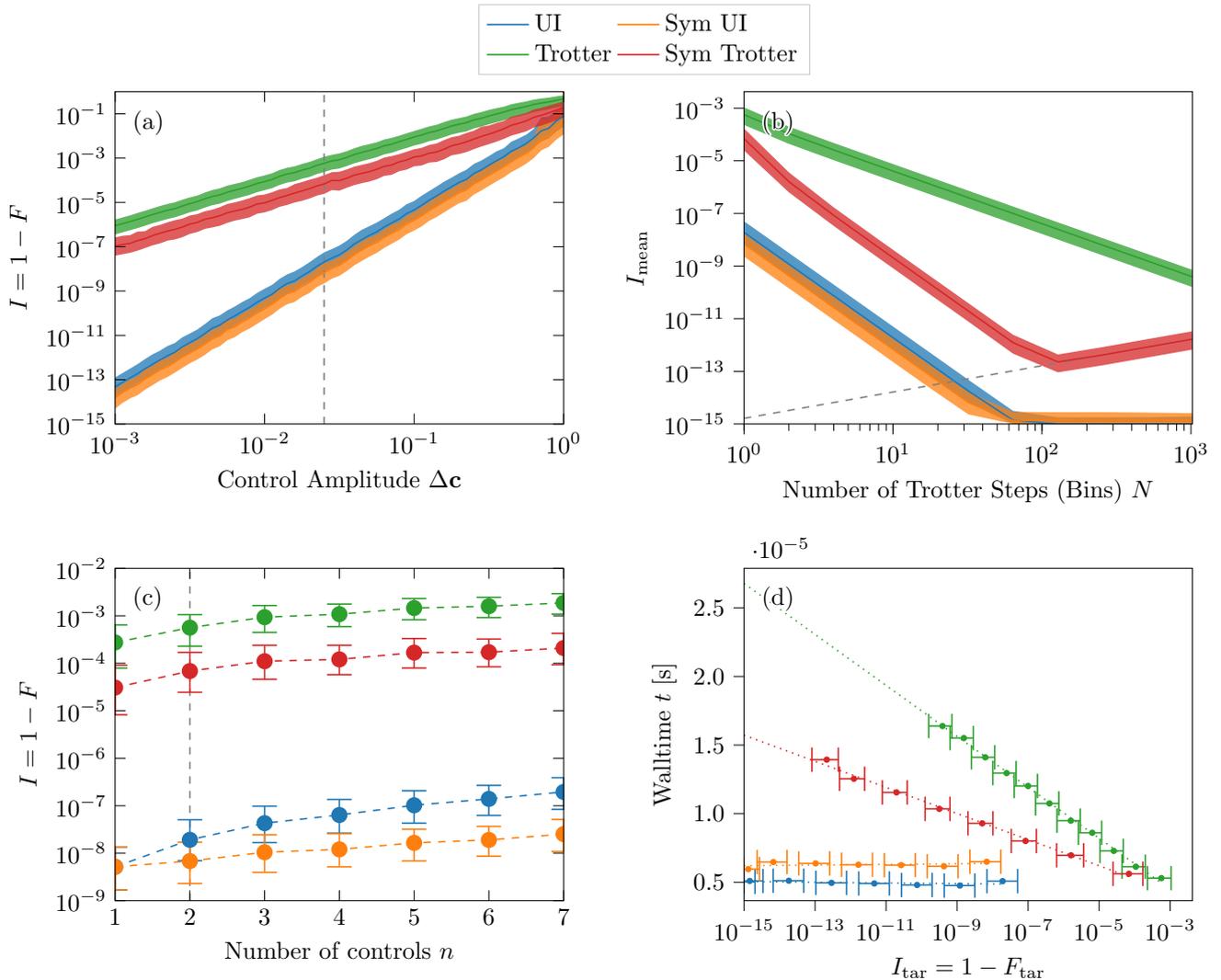

    \centering
    \includepdf{Infidelity}{infidelities_and_speed_combined}
    \caption{ Comparison of (symmetric) UI, and (symmetric) Trotterisation methods for construction of unitaries.
        Infidelity $I$ is shown as a function of (a) the maximum parameter amplitudes $c_{p, \mathrm{max}}$,
        (b) the number of Trotter steps or bins $N$ and (c) the number of system parameters $n$.
        The time $t$ to achieve a target Infidelity $I_\mathrm{tar}$ is shown in (d).
        Results are sampled 100 times for every datapoint from $d=16$ dimensional random Hamiltonians,
        constructed via the procedure in \ref{sec:random_hamiltonians},
        with $\std(\bE_p)=\nicefrac{\pi}{2} \, \forall p \in  \{1,\dots,n\}$.
        By default the non varied parameters are set to the maximal parameter amplitude of $c_{p, \mathrm{max}} = 0.025$,
        $n=2$ system parameters and a single $N=1$ Trotter step (interpolation bin),
        these default values are highlighted by the vertical gray dashed lines when varied in a plot.
        \label{fig:ui_trotter_comparison_along_dimensions}}
\end{figure*}

We compare the behaviour of the UI and its symmetrisation (Sym UI), which is derived in App.~\ref{sec:sym_ui_bch},
with that of Trotterisation (Trotter) and Symmetric Trotterisation (Sym Trotter).
Corresponding numerical results are shown in Fig.~\ref{fig:ui_trotter_comparison_along_dimensions}.

The infidelities in (a) and (b) converge according to the predictions in Tab.~\ref{tab:convergence_orders_by_method}.
In (b) we can see that the UI, Sym UI  and Symmetric Trotterisation are shifted by a constant factor in the number of steps (bins) $N$.
This factor strongly depends on the parameter amplitudes $c_{p, \maxi}$ (with $c_{p,\mini}=0$).
In (a) we show the infidelity as a function of the parameter amplitudes $c_{p, \maxi}$,
highlighting the quadratic improvement of the UI methods over the Trotterisations, for smaller parameter amplitudes.
The Sym. UI improves on the UI only by a constant prefactor, which in our examples led to a $4$ fold reduction in the infidelity,
which will generally not make up for the additional computational cost.

In (c) the infidelity of a single step (bin) is shown as a function of the number of system parameters $n$.
Finally in (d) we show the wall-time required to achieve a target infidelity $I_\mathrm{tar}$ for a 16 dimensional Hilbert space.
The UI can achieve machine precision (d), without incurring additional computational costs beyond the initial caching and the computations of a single UI step.
Trotterisation on the other hand accumulates initial numerical errors (gray dashed line) through the repeated squaring of the initial step in the computation of $N=2^j$ steps.
Furthermore the repeated squaring of (Sym.) Trotter steps add to the computational cost (d).
We find an extrapolated speedup of $3$ times for the UI over the symmetrized Trotterisation in the construction of unitaries,
when extrapolating the fidelity improvements of the Trotterisation methods. Importantly, as was shown in subplot (b), such fidelities cannot be achieved though.

\begin{figure*}[tb]
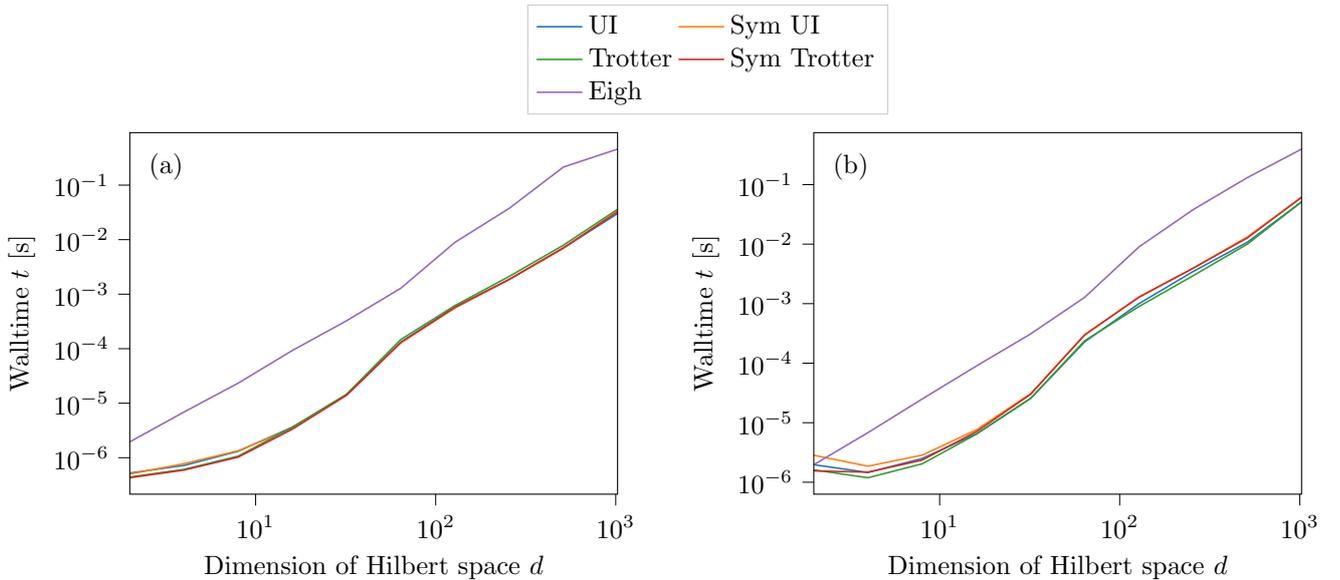

    \centering
    \includepdf{Speed}{times_by_dim_combined}
    \caption{Walltime to compute a unitary as a function of the Hilbert space dimension $d$. We compare Eigh, a single (Sym.) Trotter step, or the (Sym.)
        Unitary Interpolation for (a) single parameter and (b) two-parameter problems.
        Both Eigh and (Sym.) Unitary Interpolation achieve machine precision in this time. } \label{fig:time_by_dim}
\end{figure*}

In Fig.~\ref{fig:time_by_dim} we compare the computational time as a function of the Hilbert space dimension for (symmetric) UI and
(Sym.) Trotter (using a single step), with (hermitian) eigenvalue decomposition based exponentiation, see App.~\ref{sec:expm_eigh}.
All methods show roughly a $\mathcal{O}(d^3)$ complexity, with roughly a 10 fold speedup for the matrix product approaches.
Greater speed-ups are expected for GPU implementations.
For a single step the Trotter approach and UI require similar wall-times, as they require the same number of matrix multiplications.

\subsection{Construction from Cached Matrices}
The interpolation in Eq.~\eqref{eq:nd_ui} can be computed from cached matrices.
In analogy to Eq.~\eqref{eq:ui_1d_product} we can decompose the one-parameter displacement operators via Shur decomposition into
eigenvector and diagonal eigenvalue matrices $\hat{V}_{\bi, \bi\pm\be_p}$ and $\bE_{\bi, \bi\pm\be_p}$. The decomposition then reads
\begin{equation}
    \left( \hU_{\bi + \bs_p} \hU_\bi^\dagger\right) = \hat{V}_{\bi, \bs_p} \expi{\bE_{\bi,\bs_p}} \hat{V}_{\bi, \bs_p}^\dagger. \label{eq:nd_displacement_operator_eigendecomposition}
\end{equation}
The diagonal eigenvector matrices $\hE_{\bi, \bs_p}$ are stored as vectors. From the eigenvector matrices we define the cached matrices
\begin{gather}
    \hat{L}_{\bi, \bs_n} = \hat{V}_{\bi, \bs_n}, \quad \hat{R}_{\bi, \bs_1} = \hat{V}_{\bi, \bs_1}^\dagger \hU_\bi,
    \hC_{\bi, \bs_{m+1}, \bs_{m}} = \hV_{\bi, \bs_{m+1}}^\dagger \hV_{\bi, \bs_{m}}. \label{eq:nd_ui_cached_matrices}
\end{gather}
Using these cached matrices, we can construct the UI via
\begin{align}
    \hU_\UI(\balpha) = & \hat{L}_{\bi+\bs_n} \expi{\bE_{\bi+\bs_n} |\alpha_n|} \cdot \nonumber                                                                                       \\
    \cdot              & \left(\prod_{p=1}^{n-1} \hat{C}_{\bi, \bs_{p+1},\bs_{p}} \expi{\bE_{\bi+\bs_p} |\alpha_p|} \right) \hat{R}_{\bi+\bs_1}. \label{eq:nd_ui_cached_computation}
\end{align}
The multiplication of the diagonal exponentials $\expi{\bE_{\bi+\bs_p} |\alpha_p|}$ can again be performed as a row-wise scaling operation, with negligible computational cost,
effectively reducing the computational cost to $n$ matrix multiplications.

\subsection{Gradients}
In \ref{sec:1d_gradients} we discussed the utility of calculating the gradients of the unitaries with respect to the system parameters, for gradient based optimization.
These gradients are efficiently calculated from the cached matrices. For the derivative $\frac{\partial \phantom{\hH}}{\partial c_p}$, we
replace the energy exponential $\exp(-i\bE_{\bi+\bs_p} |\alpha_p|)$ in Eq.~\eqref{eq:nd_ui_cached_computation}, with its derivative
$-\frac{i\bE_{\bi+\bs_p} N_p}{\Delta c_p} \exp(-i\bE_{\bi+\bs_p} |\alpha_p|)$.

To avoid unnecessary matrix operations in the computation of the derivatives, we adapt the differentiation trick from \cite{Khaneja2005}.
Let $\hU(\bc)=(\prod_{p=1}^{n}  \hC_p \hA_p(\alpha_p)) \hC_0$ be the matrix product of an $n$-parameter interpolation, where we neglected the grid indexes for readability, replaced $\hat{L}=\hC_n$, $\hat{R}=\hC_0$, and the diagonal-matrix exponential terms with $\hA_p(\alpha_p)$.
The gradients can be calculated efficiently via the chain rule, by first definining $\hat{P}_0 = \hU(\bc)\hA_0^\dagger(c_0)\hC_0^\dagger$,
$\hat{Q}_0 = \hC_0$ and then iteratively starting from $i=1$ to $i=N$ calculating $\hat{P}_i = \hat{P}_{i-1}(\hA_i(c_i)\hC_i)^\dagger $, then $\ddfrac{\hU(\bc)}{c_i}=\hat{P}_i \ddfrac{\hA_i}{c_i} \hat{Q}_{i-1}$ and then updating $\hat{Q}_i = (\hC_i \hA_i(c_i)) \hat{Q}_{i-1}$.

This requires only 3 additional matrix multiplications per derivative, and compares favorably with diagonalisation based differentiation,
which requires 4 additional matrix multiplications per system parameter, see Eq.~(10) in \cite{Dalgaard2020}.

\begin{figure*}[tb]
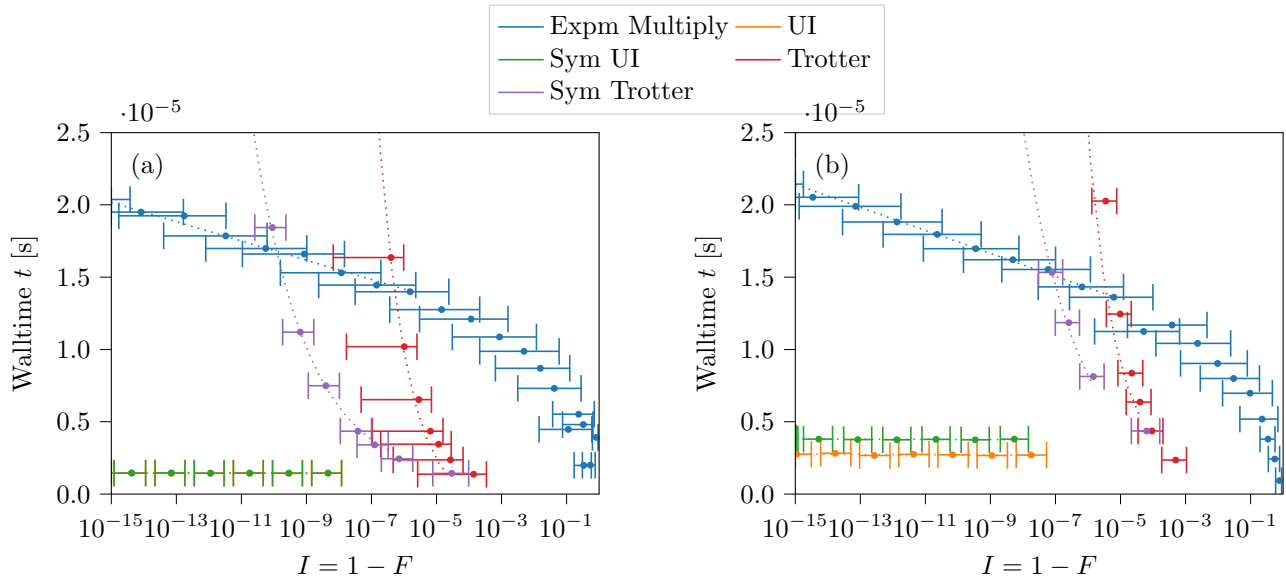

    \centering
    \includepdf{Speed}{time2infidelity_vectors}
    \caption{Walltime to infidelity at system amplitude ratio 0.01 for 15 dimensional wavefunctions for (a) single parametrer and (b) two-parameter problems.
        The Trotter based methods perform worse, as every step has to be executed.
        Expm Multiply computes the action of the Hamiltonian onto the wavefunction using the scaling and squaring algorithm.
        We find the Unitary Interpolation to be about 10 times as fast as calculating the action of the Hamiltonian via scaling and squaring (Krylov).\label{fig:time_by_infidelity_vectors}}
\end{figure*}

\section{State Evolution}\label{sec:state-evolution}
Many problems, such as state transfer optimizations \cite{Khaneja2005} or optimizations of subsystem dynamics \cite{motzoi2011},
do not require the full unitary, but only the action of the unitary onto a wavefunction $\ket{\psi} = \hU \ket{\psi_0}$ or a (sampled) set of wavefunctions.

This allows us to replace matrix-matrix operations $\mathcal{O}(d^3)$ with matrix-vector operations $\mathcal{O}(d^2)$,
and presents computational advantages in the case of the UI. For Trotterisations  on the other hand,
the repeated squaring of a single Trotter step, to create exponentially many Trotter steps
($2^j$ Trotter steps via $j$ matrix multiplications) is no longer possible, and each Trotter step has to be performed sequentially.
The divide and conquer method used in the eigendecomposition of the Hamiltonian can not profit from this, and is replaced by
direct computation of the action of the matrix exponential onto the wavefunction via scaled Taylor expansion into the wavefunction \cite{AlMohy2011}.

In order to compare this approach with the other methods,
we (reverse engineered and) Cythonized \cite{behnel2011cython} the \textit{scipy} implementation \cite{2020SciPy} (\lstinline{scipy.sparse.linalg.expm_multiply}).
We reuse the computation of the hyperparameters to reach a target accuracy between different exponentiations further speeding up the procedure
with respect to the scipy implementation.

In Fig.~\ref{fig:time_by_infidelity_vectors} we compare the time to reach a target infidelity for the different methods. The UI achieves machine precision
one order of magnitude faster than the calculation of the action of the exponential, which furthermore outperforms Trotter based methods.

\section{Caching}\label{sec:caching}
The UI replaces the repeated diagonalisation of Hamiltonians with the initial creation of an interpolation cache.
This cache scales on the order of $\bigO(\prod_{i=1}^{n} N_i)$,
where $N_i$ is the number of interpolation bins along the $i$-th parameter direction
(a precise formula for the cache size is derived in App.~\ref{sec:cache_size}).

In order to keep the cache requirements low, we optimize the cache size while keeping below a target infidelity $I_\mathrm{tar}$.
In App. \ref{sec:optimal_caching}, we first derive a method to estimate the maximum infidelity of a particular binning $\mathbf{N}$,
as a polynomial function of values derived from a test binning.
We then use this polynomial to optimize the binning $\mathbf{N}$ in a three step process, to determine an optimum within milliseconds.
We compare the quality of the bin optimisation with results from exhaustive search, in Tab.~\ref{tab:acc_opt_cache},
sampled 100 times for different parameter amplitude combinations, $c_{p, \maxi} \in {0.25,1.0}$ for Hamiltonians with $\mathrm{std}(\hat{E}_p)=\nicefrac{\pi}{2}$, and number of parameters $n$.
This demonstrates that we can estimate the infidelity with $< 10\%$ error, and find bin optima typically within
$< 1\%$ of the optimal binning. 

\begin{table*}[t]
    \centering
    \caption{Required Cache size $C$ as a function of system parameters $n$, the maximum parameter amplitudes $c_{p,\mathrm{max}}$ (and $c_{p,\mathrm{min}}=0$),
        the error of the optimisation in determining the optimal binning, and the accuracy of the infidelity estimation.
        Cache is supposed to achieve $I_\mathrm{tar} = 10^{-12}$. The worst estimation with a relative Error $> 1\%$ is highlighted in red. \label{tab:acc_opt_cache}}
    \begin{tabular}{c c c c c}\hline
        n   & Parameter Amplitudes                  & Cache Size        & Rel. Err. Cache Optimum                  & Rel. Acc. Infidelity      \\
        \hline
        $1$ & $[2.5]\num{e-2}$                      & $9.6 \pm 1.7$     & $0 \pm 0$                                & $(1.7 \pm 1.3)\num{e-3}$  \\
        $2$ & $[2.5,\, 1]\num{e-2}$                 & $(1 \pm 0.3)e2$   & $(1.9 \pm 6.8)\num{e-3}$                 & $(8.7 \pm 8)\num{e-2}$    \\
        $2$ & $[2.5,\, 2.5]\num{e-2}$               & $(2.4 \pm 0.7)e2$ & $(9.3 \pm 44.1)\num{e-4}$                & $(2.10 \pm 2.6)\num{e-2}$ \\
        $3$ & $[2.5,\, 1,\, 1]\num{e-2}$            & $(1.3 \pm 0.5)e3$ & $(4.6 \pm 12.3)\num{e-3}$                & $(8.9 \pm 6.5)\num{e-2}$  \\
        $3$ & $[2.5,\, 2.5,\, 2.5]\num{e-2}$        & $(7.8 \pm 2.6)e3$ & $(1.1 \pm 3.7)\num{e-3}$                 & $(7.4 \pm 5.4)\num{e-2}$  \\
        $4$ & $[2.5,\, 1,\, 1,\, 1]\num{e-2}$       & $(2 \pm 0.9)e4$   & $\color{red} (?) (2.6 \pm 4.3)\num{e-2}$ & $(8.10 \pm 7.6)\num{e-2}$ \\
        $4$ & $[2.5,\, 2.5,\, 2.5,\, 2.5]\num{e-2}$ & $(3.1 \pm 1.3)e5$ & $(5.8 \pm 12.4)\num{e-3}$                & $(9.4 \pm 6.4)\num{e-2}$  \\
        \hline
    \end{tabular}
\end{table*}

\section{Conclusion}
In this work we derived a new framework for the repeated computation of unitaries $\hU(\bc)$ generated by parametric Hamiltonians $\hH(\bc)$ with varying system parameters $\bc$. 
We used two key insights to speed up such computations, first that interpolation guarantees fourth order convergence
in system parameters $c_p$ and interpolation bins $N_i$,
and second that the interpolation can be performed on a grid, which allows us to (in principle) reach arbitrary precision.
The interpolation requires the same amount of matrix multiplications as a single Trotter step, but with the benefit of only requiring one step to achieve machine precision,
whereas Trotterisation needs to be repeated to achieve our desired target accuracies.

We optimized the binning to furthermore reduce the time and space needed for caching matrices as this
remains the main bottleneck especially for systems with many system parameters or large variation in the parameter amplitudes $\Delta c_p = c_{p,\mathrm{max}} - c_{p,\mathrm{min}}$.
A further limitation is the reliance on dense operators; we have not investigated whether the interpolation can bring additional speed-up for sparse operators.

We derived the properties of the UI via the properties of the introduced displacement operators and the roots of the interpolation.
By adding more roots to the interpolation, the order of convergence can be further improved, which we leave for future work.

\begin{acknowledgments}
    This work was funded by the Federal Ministry of Education and
    Research (BMBF) within the framework programme "Quantum technologies – from
    basic research to market" (Project QSolid, Grant No.~13N16149, Project Spinning, Grant No.~13N16210),
    by the Deutsche Forschungsgemeinschaft (DFG, German Research Foundation) under Germany’s
    Excellence Strategy – Cluster of Excellence Matter and
    Light for Quantum Computing (ML4Q) EXC 2004/1 –
    390534769 and
    by HORIZON-CL4-2022-QUANTUM-01-SGA Project under Grant 101113946 OpenSuperQPlus100.
\end{acknowledgments}

\section*{Implementation}
All methods were implemented in Cython (C compiled for use in python) \cite{behnel2011cython} and are available at \url{https://github.com/Ntropic/unipolator}.
The code can be installed via pip using \lstinline{pip install unipolator}.


\bibliography{bibliography}

\onecolumngrid
\appendix

\section*{Symbols}
 {\renewcommand{\arraystretch}{1.5}
  \begin{tabularx}{\linewidth}{lX} 
      $n$                          & The number of system parameters                                                                                                                                                \\
      $N$                          & The number of bins                                                                                                                                                             \\
      $d$                          & The Hilbert space dimension of the quantum system(s)                                                                                                                           \\
      $m$                          & The number of time steps                                                                                                                                                       \\
      $\hat{X}$                    & A complex square matrix, eg. $\hH$, $\hU$ and $\hV$ for Hamiltonian, unitary operators and eigenvectors respectively                                                           \\
      $\mathbf{X}$                 & A complex numbered vector, eg. $\bE$ for the eigenvectors of a schur or eigh decomposition                                                                                     \\
      $\mathbf{x}$                 & An integer valued vector, eg. $\bi$ for the indices of the eigenvectors of a schur or eigh decomposition                                                                       \\
      $\balpha$                    & A real valued vector, eg. $\balpha$ for the interpolation coefficients                                                                                                         \\
      $\prod_{i=1}^{n}$            & The product from the left side, eg. $\prod_{i=1}^s x_i = x_s \cdots x_2 x_1$, as opposed to the product from right to the left side $\coprod_{i=1}^{n}$                        \\
      $\bE,\, \hV = \eig(\hat{X})$ & An eigenvalue decomposition, using  \lstinline{eigh} (Divide and Conquer) and \lstinline{shur} (for Shur decomposition) for hermitian and non-hermitian matrices respectively. \\
      $\expi{\bE_j \alpha}$        & A diagonal matrix constructed from the element wise exponentials $\diag{\expi{E_{n,1} \alpha}, \cdots, \expi{E_{j,N} \alpha}}$
  \end{tabularx}}

\section{Finding the Associated Voronoi Cell} \label{sec:voronoi_cell}
In order to determine the interpolation cell, in which a point $\bc$ lies,
we need to find the closest odd-summed lattice-site $\bi_c$.
We do this by first finding the closest lattice-site $\tilde{\bi}=(\tilde{i}_1, \dots, \tilde{i}_n)$
\begin{equation}
    \tilde{i}_p = \left\lfloor \frac{N (c_p - c_{p,\mathrm{min}}) + 0.5}{\Delta c_p} \right\rfloor 
\end{equation}
If the lattice-site is odd-summed $\sum_{p=1}^n \tilde{i}_p \mod 2 = 1$, then $\bi_c = \tilde{\bi}$, otherwise we need to find the closest odd-summed lattice-site $\bi$.
To do so, we calculate the relative location $\tilde{\balpha}$ between $\bc$ and the closest lattice-site
\begin{equation}
    \tilde{\balpha} = \bc - \bc_{\tilde{\bi}}, \quad \mathrm{where} \quad \bc_{\tilde{\bi}} = \bc_{\mathrm{min}} + \tilde{\bi} \frac{\Delta \bc}{\bN}. \label{eq:nd_lattice_site_relative}
\end{equation}
The interpolation cell $\bi$ is then found by moving in the direction of the largest component $ m = \mathrm{argmax}_p \left|\tilde{\alpha}_p\right|$ of $\tilde{\balpha}$,
so that
\begin{equation}
    \bi = \tilde{\bi} + \mathrm{sign}(\tilde{\alpha}_m) \be_m. \label{eq:nd_lattice_site_odd}
\end{equation}
In case there are multiple largest components, we choose the one with the smallest index $m$.

\section{Asymmetric Standard deviations} 
\begin{figure}[t]
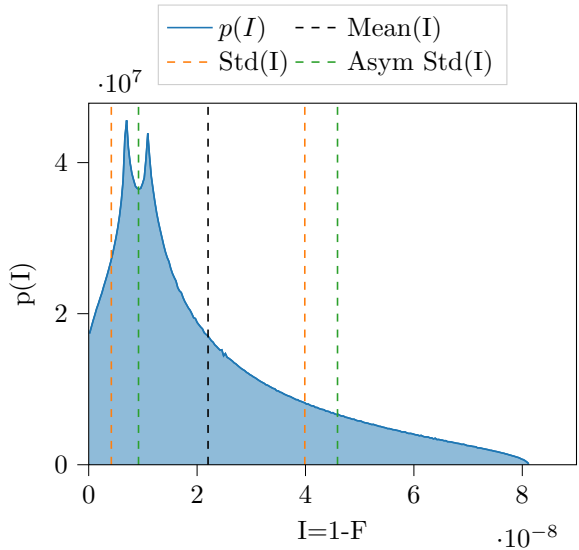

    \centering
    \includepdf{Fitting}{infidelity_distribution_mean_std_hist_2d_UI}
    \caption{The infidelity distribution of a 2 dimensional UI,
        its mean, and the bounds of both the classical and asymmetric standard deviation.
        The values below the mean are clustered more strongly in 2 peaks, wheras the values above the mean are more spread out,
        with very few large values contributing the most to the standard deviation. \label{fig:inf_dist_2d}}
\end{figure}
The infidelity distribution of the UI (but also of the Trotterisation shemes) as a function of the system parameters is not normally distributed, see Fig.~\ref{fig:inf_dist_2d}.
As a result the use of a standard deiation can be misleading.
Furthermore as the average infidelity approaches zero (with increasing numbers of bins or Trotter steps) the standard deviation can be larger than the average,
which would suggest values outside of the possible range $I \in[0,1]$.
We therefore decided to use an asymmetric standard deviation, with separately calculated standard deviations for values below and above the average infidelity.
We calculated this standard deviation using the following formula
\begin{equation}
    \begin{aligned}
        \std_\mini (\mathbf{I}) = & \sqrt{\frac{\sum_{I_i \leq I_\mathrm{avg}} (I_i - I_\mathrm{avg})^2}{(\sum_{I \leq I_\mathrm{avg}} 1) -1}} \\
        \std_\maxi (\mathbf{I}) = & \sqrt{\frac{\sum_{I_i \geq I_\mathrm{avg}} (I_i - I_\mathrm{avg})^2}{(\sum_{I \geq I_\mathrm{avg}} 1) -1}}
    \end{aligned}
\end{equation}
The asymmetric standard deviations are used throughout this paper and either given by error bars or by a transparent area around a line plot.

\section{Estimating Average Infidelities} \label{sec:average_infidelity}
In this paper we construct the average infidelity of the interpolation (approximation) with respect to the exact unitary. For this purpose we average the infidelity over
\begin{enumerate}
    \item All possible states (analytically),
    \item Sampled Hamiltonians with a gaussian eigenvalue distribution (numerically),
    \item The parameters within the interpolation/approximation hypervolume (semi-analytical).
\end{enumerate}
In the following sections we will explain the procedures for each of those averaging procedures. For completeness we also estimate the maximum infidelity over the space of interpolation/approximation parameters.

\subsection{Averaging Infidelity over Possible States}
The Fidelity $F(\ket{\Phi})$ for an initial state $\Phi$, between an approximated time evolution $\hat{U}_\mathrm{ap}$ and the exact evolution $\hat{U}_\mathrm{ex}$ is given by the probability of the approximated final state $\hat{U}_\mathrm{ap}\ket{\Phi}$ to collapse onto the exact final state $\hat{U}_\mathrm{ex}\ket{\Phi}$ when measured in a measuring basis in which the final state is an eigenstate.
\begin{align}
    F(\ket{\Phi})=\left|\braket{\Phi|\hat{U}_{\mathrm{exact}}^\dagger\hat{U}_{\mathrm{approx}}|\Phi}\right|^2.
\end{align}
Using the overlap operator $\hat{M}=\hat{U}_{\mathrm{exact}}^{\dagger}\hat{U}_{\mathrm{approx}}$ the Fidelity can be written in the short form
\begin{align}
    F(\ket{\Phi})=\left|\braket{\Phi|\hat{M}|\Phi}\right|^2.
\end{align}
Equivalently the infidelity is defined as $I(\ket{\Phi})=1-F(\ket{\Phi})$, and corresponds to the rate of error for a measurement in the eigenbasis of $\hat{U}_{\mathrm{exact}}$.
Integrating the infidelity $F(\ket{\Phi})$ over the space of all normalized wavevectors $\ket{\Phi}$ we use
\begin{gather}
    \begin{aligned}
        I_\mathrm{av}\stackrel{\mathrm{def}}{=} & 1-\int\vert\bra{\Phi}\hat{M}\ket{\Phi}\vert^2 dS^{2d-2}                                      \\
        =                                       & \frac{d}{d+1} -\frac{\vert\mathrm{Tr}(\hat{M})\vert^2}{d(d+1)},\label{eq:average_infidelity}
    \end{aligned}
\end{gather}
for the average infidelity, as derived in 4.4 in \cite{Dankert2005} and more approachably rederived in \cite{Pedersen2007}.
where $d$ refers to the $d$ dimensions of the overlap operator $\hat{M}\in \mathds{C}^{d\times d}$.

\subsection{Averaging Infidelity over Random Hamiltonians} \label{sec:random_hamiltonians}
We average the infidelities using random Hamiltonians $\hH$ of Hilbert space dimension $d$ with normally distributed random eigenvalues $\bE_i = \mathcal{N}(0, \sigma^2)$, as this emulates the eigenvalue distribution of large collections of particles (Stirling limit).
Starting from these randomly selected eigenvalues, we construct the Hamiltonians  via the transformation $\hH = \hU \mathrm{diag} (\bE) \hU^\dagger$,
where $\hU$ is a random unitary matrix (generated from a Haar measure).
\subsubsection{Generating Haar measure random unitaries}
We generate $s$ dimensional random unitaries $\hU$ (from a Haar measure) using the following procedure,
\begin{enumerate}
    \item Generate a random complex matrix $\hat{X}$ of size $d \times d$ with real and imaginary parts chosen using normally distributed pseudo random numbers (using the PCG-64 algorithm) $\hat{X}_{i,j} = \mathcal{N}(0,1)+i\mathcal{N}(0,1)$,
    \item Extract a unitary matrix $\hU$ from $\hat{X}$ using the QR decomposition (Gram-Schmidt orthogonalization and normalization)
          $\hU, \hat{R} = \mathrm{QR}(\hat{X})$.
\end{enumerate}

\subsection{Averaging Infidelity over the Interpolation/Approximation Volume}
We estimate the average infidelity over the interpolation/approximation by fitting a set of basis functions $u_\UI(\balpha, \bj)$ to samples of the infidelity within the interpolation/approximation volume.
For each basis function we cache the average infidelity, which we calculate analytically.
The average infidelity is then given by the weighted sum of the basis function averages, where the weights are the coefficients of the basis function expansion, determined via the regression procedure.

\subsubsection{Basis Functions}
We choose n-dimensional polynomial basis functions of degrees $1$ to $m$,
\begin{equation}
    u(\balpha,\bj) = \prod_{i=1}^{n} \alpha_i^{j_i} \with 1 \leq\sum_{i=1}^{n} j_i \leq m, \label{eq:basis_nd}
\end{equation}
using the fact that at $\balpha = 0$ both approximation and interpolation is exact ($I_\mathrm{av}(\balpha=0)=0$). We then approximate the infidelity as a linear combination of the basis functions
\begin{equation}
    I(\balpha) \approx \sum_{\bj\in \mathds{N}^{n}}^{1\leq \norm{\bj} \leq m} c_\bj u(\balpha, \bj). \label{eq:I_basis_expansion}
\end{equation}
The coefficients $c_\bj$ are estimated by linear regression of sample values of the infidelity at randomly chosen points in the interpolation volume.
There are $\binom{m+n}{n}-1$ basis functions of degrees $1\leq \norm{\bj} \leq m$, which we can estimate accurately by sampling $\binom{m+n}{n}-1$ points. For UI, we can furthermore use the fact,
that along the axis the infidelity returns to zero at the corner $I(\balpha = \hat{e}_k) = 0$, where $\hat{e}_k$ is the unit vector in the $k$-th direction. This reduces the number of required sample points by $n$.

\subsubsection{Average Infidelities}
We calculate the average infidelity by integrating over the interpolation/approximation volume. This is accomplished by calculating and caching the average infidelity of the basis functions $\overline{u}(\bj)$ over the interpolation volume, and reusing the regression coefficients $c_\bj$
\begin{equation}
    I_\mathrm{av} \approx \sum_{\bj\in \mathds{N}^{n}}^{1\leq \norm{\bj} \leq m} c_\bj \overline{u}(\bj). \label{eq:I_av_basis_expansion}
\end{equation}
The integration of the basis functions has to be performed separately for UI and Trotter techniques, due to the different interpolation volumes. We focus on the positive quadrant $\alpha_i \leq 0 \, \forall i\in[1,n]$, as all other quadrants are structurally equivalent.

\paragraph{Average Infidelity of Trotterisations}
For Trotterisations the interpolation volume is given by a hypercube, allowing for a separation of the n-dimensions in the integral
\begin{align}
    \overline{u}_{\Tr}(\bj) & = \frac{1}{\mathcal{V}_{\Tr}} \int_{\mathcal{V}_{\Tr}}  u_{\Tr}(\balpha, \bj) \td\balpha \nonumber                         \\
                            & = \prod_{i=1}^{n} \int_0^1 \alpha_i^{j_i} \td \alpha_i = \prod_{i=1}^{n} \frac{1}{j_i+1}. \label{eq:average_infidelity_tr}
\end{align}

\paragraph{Average Infidelity of the Unitary Interpolation}
In the case of the UI the solution is not as straightforward, as we need to integrate over the interpolation cells.
Using Eq.~\eqref{eq:control_param2interpolation_param_nd}, the definition of the Voronoi cells, Eq.~\eqref{eq:voronoi_cell}, can be restated as
\begin{equation}
    \begin{gathered}
        \alpha_p \in (-1,1) \quad \forall \quad p \in \{1,\dots,n\}, \\
        |\alpha_p| \geq 1- |\alpha_q| \quad \forall \quad q \neq p \in \{1,\dots,n\}. \label{eq:interpolation_volume_condition_2}
    \end{gathered}
\end{equation}
For every direction the integral is limited by the largest component, therefore we split the integral into $n$ cases, for the $n$ possible largest components in $\balpha$.
\begin{align}
    \overline{u}_{\UI}(\bj) & = \frac{1}{\mathcal{V}_{\UI}}\int_{\mathcal{V}_{\UI}} u(\balpha, \bj) \td\balpha \nonumber                                                                                             \\
                            & = \frac{1}{\mathcal{V}_{\UI}} \sum_{i=1}^{n} \int_0^1 \alpha_i^{j_i}  \left(\prod_{h\neq i}^{n} \int_0^{\min(\alpha_i, 1-\alpha_i)}  \alpha_h^{j_h}\td \alpha_h \right)  \td \alpha_i,
\end{align}

with the hypervolume $\mathcal{V}_{\UI} = 2^{n-1}$. This integral further splits up into the two cases $\alpha_1 \in [0,\nicefrac{1}{2}]$ and $\alpha_1 \in (\nicefrac{1}{2},1]$, to avoid the $\min$ function in the integral bounds.
\begin{align}
    \overline{u}_{\UI}(\bj) & = \frac{1}{\mathcal{V}_{\UI}}\sum_{i=1}^{n} \int_0^{\nicefrac{1}{2}} \alpha_i^{j_i} \left( \prod_{h\neq i}^{n} \int_0^{\alpha_i} \alpha_h^{j_h}\td \alpha_h \right) \td \alpha_i + \frac{1}{\mathcal{V}_{\UI}}\sum_{i=1}^{n} \int_{\nicefrac{1}{2}}^1 \alpha_i^{j_i} \left( \prod_{h\neq i}^{n} \int_0^{1-\alpha_i} \alpha_h^{j_h}  \td \alpha_h \right) \td \alpha_i  \nonumber \\
                            & = \frac{1}{\mathcal{V}_{\UI}}\sum_{i=1}^{n} \int_0^{\nicefrac{1}{2}} \alpha_i^{j_i} \left( \prod_{h\neq i}^{n} \frac{\alpha_i^{j_h+1}}{j_h+1} \right) \td \alpha_i + \frac{1}{\mathcal{V}_{\UI}}\sum_{i=1}^{n} \int_{\nicefrac{1}{2}}^1 \alpha_i^{j_i} \left( \prod_{h\neq i}^{n} \frac{(1-\alpha_i)^{j_h+1}}{j_h+1} \right) \td \alpha_i.
\end{align}

This can be expanded into a polynomial sum and can be evaluated quickly for every basis function (in our case via vector operations).
\subsubsection{Standard deviations}
In order to evaluate the asymmetric standard deviations from the average infidelity, we sample the fitted infidelity function at $100$ randomly chosen sample points. This allows a fast evaluation of the standard deviation
as the contribution of every basis function can be cached for every sample point, avoiding the evaluation of the matrix exponentials.
\subsubsection{Quality of the basis function estimates}
\begin{figure*}
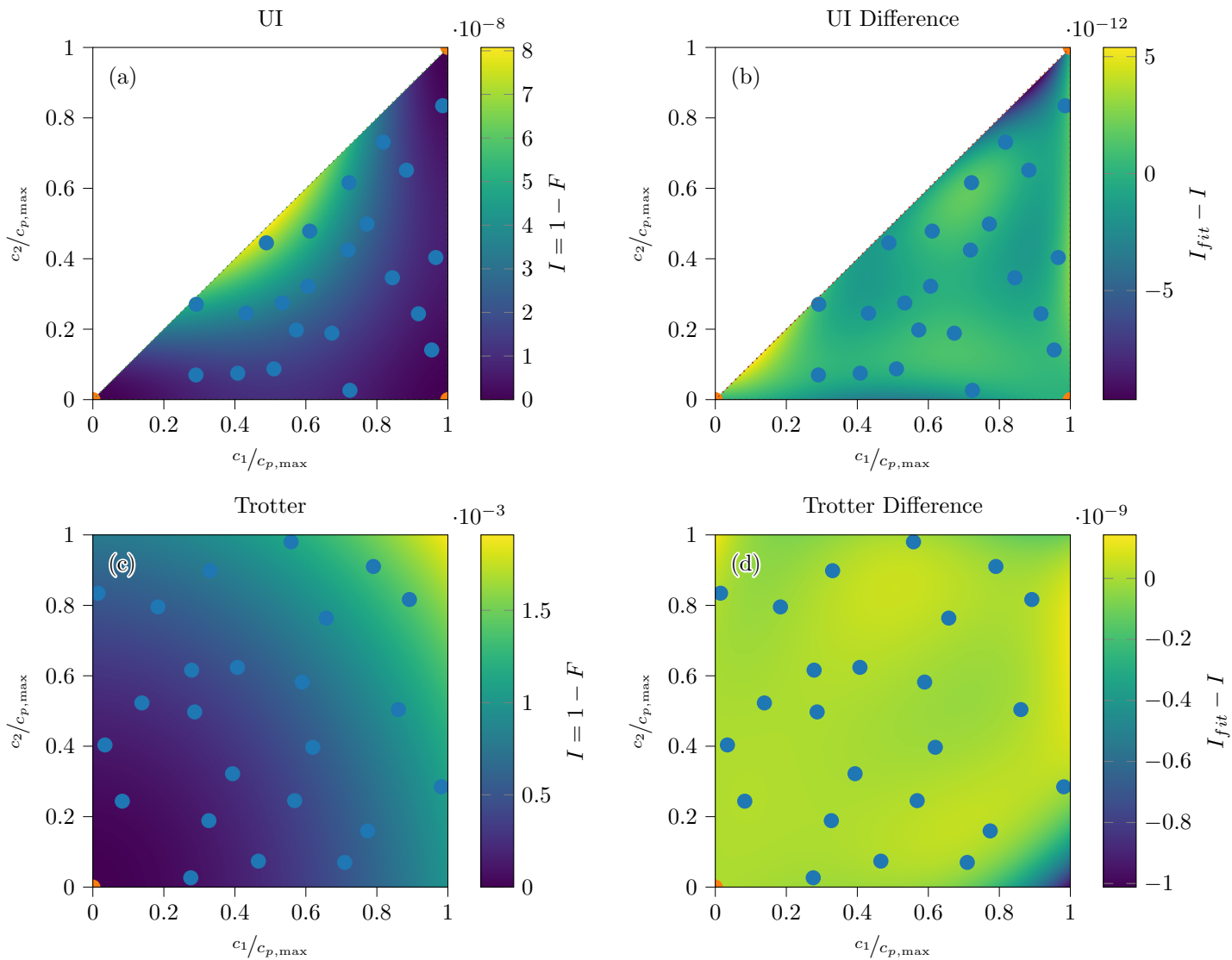

    \centering
    \includepdf{Fitting}{fit_2d_combined}
    \caption{Infidelity maps of 2 parameter UI (a) and Trotterisation (c) with sample points at a point ratio of 2.0 in \textcolor{blue}{blue} and boundary condition points in \textcolor{orange}{orange} with known infidelity of $0.0$,
        from which we construct an approximation to the infidelity map using 2nd order Taylor expansion (neglecting 0th order terms).
        The difference between the estimation and the true infidelity map for Unitary Interpolation and Trotterisation are shown in (b) and (d) repsectively. We find a relative accuracy of $<10^{-4}$ for the Unitary Interpolation and $<10^{-7}$ for the Trotterisation.
        The fit works extremely well close to the sample points used in the estimation, the decrease in accuracy can be seen in the top-right corner of (b) and bottom-right corner of (d). \label{fig:fit_2d_combined}}
\end{figure*}

\begin{figure}[t]
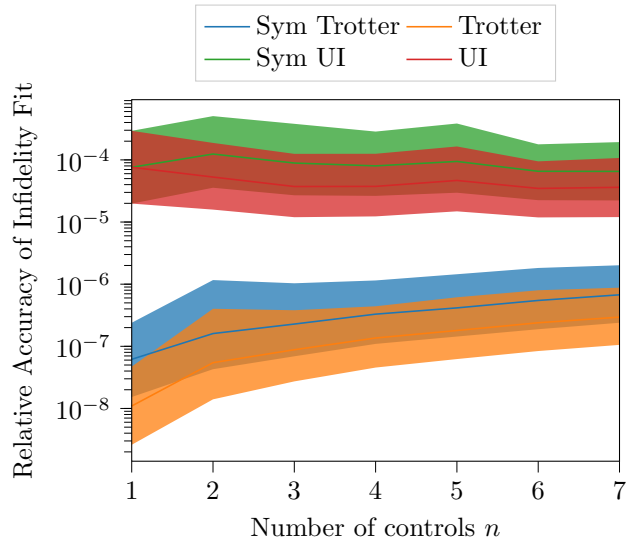

    \centering
    \includepdf{Fitting}{rel_acc_infidelity_fit}
    \caption{The relative accuracy of the fit using a point ratio of $2.0$, for Trotterisation and Unitary Intepolation techniques as a function of the number of system parameters.
    The relative accuracy is computed for 100 systems with 100 sample points each. At each of the the sample points $i \in \{1,\dots,100\}$ we calculate the difference
    $d_i=|I_{i, \mathrm{fit}} - I_{i, \mathrm{true}}|$ between the true infidelity at the point and the prediction prediction using the fit technique.
    The relative accuracy at each point is calculated via $r_i = \frac{d_i}{\mathrm{mean}(I_\mathrm{true})}$. Importantly we find relative accuracies beyond $10^{-3}$ and higher accuracy for less accurate methods.\label{fig:relative_fit_accuracy_by_dim}}
\end{figure}

We test the quality of the basis function approach to the average infidelity using 2 times as many sample points as fit parameters.
In figure \ref{fig:fit_2d_combined} we show the infidelity maps of the 2 dimensional UI and Trotterisation and the deviations from the true infidelity as a function of the system parameters.
We find low relative errors of $<10^{-4}$ for the UI and $<10^{-7}$ for the Trotterisation across the space of system parameters with high accuracies close to the sample points used in the estimation.
We found that using just as many sample points as fit parameters is sufficient to obtain a good approximation as long as the infidelities do not approach the double precision limit of $~10^{-16}$,
we averted this problem via the choice of 2 times as many sample points as fit parameters. In figure \ref{fig:relative_fit_accuracy_by_dim} we explore the behaviour of the fit method for UI, Symmetric UI, Trotterisation and Symmetric Trotterisation as a function of number of system parameters.
For The UI based approaches the accuracy remains stable, wheras for Trotterisations it decreases from $~10^{-7}$ to $~10^{-4}$ for 2 and 7 system parameters respectively.

\begin{table}[t]
    \centering
    \caption{The accuracy of the approximation in Eq.~\eqref{eq:infidelity_bin_approximation} using randomly generated and trace-orthogonalized Hamiltonians $\hH_p$.
        The relative accuracy $\frac{I_\mathrm{est}-I_\mathrm{exact}}{I_\mathrm{exact}}$ is estimated for different numbers of parameter terms $n \in[2,3,4]$
        and Hilbert space dimension $s \in \{4,16\}$, each sampled 100 times according to Sec.~\ref{sec:random_hamiltonians}. In one and two parameter dimensions the relative error is limited by machine precision.
        For larger $n$ the relative errors are about one order of magnitude smaller than the values.
        The trace-orthogonalazied Hamiltonians behave similar to the unorthogonalized ones. \label{tab:traceless_traces}}
    \begin{tabular}{c c c c }\hline
        n   & s    & Rel. Acc.                 & Rel. Acc. (Orthogonalized) \\
        \hline
        $1$ & $4$  & $(6.59\pm49.48)\num{e-6}$ & $(1.65\pm8.79)\num{e-6}$   \\
        $1$ & $16$ & $(7.87\pm9.54)\num{e-9}$  & $(1.06\pm1.50)\num{e-8}$   \\
        $2$ & $4$  & $(3.31\pm4.82)\num{e-8}$  & $(4.67\pm8.05)\num{e-8}$   \\
        $2$ & $16$ & $(1.92\pm0.95)\num{e-8}$  & $(1.64\pm0.95)\num{e-8}$   \\
        $3$ & $4$  & $(2.71\pm2.75)\num{e-1}$  & $(2.25\pm2.24)\num{e-1}$   \\
        $3$ & $16$ & $(6.42\pm4.72)\num{e-2}$  & $(7.24\pm5.47)\num{e-2}$   \\
        $4$ & $4$  & $(3.37\pm3.04)\num{e-1}$  & $(2.81\pm2.55)\num{e-1}$   \\
        $4$ & $16$ & $(9.74\pm7.69)\num{e-2}$  & $(8.32\pm5.91)\num{e-2}$   \\
        \hline
    \end{tabular}
\end{table}

\section{(Hermitian) Eigenvalue Decomposition}\label{sec:expm_eigh}
In Fig.~\ref{fig:time_by_dim} we compare the matrix product approaches (Trotterisations and UI) to eigenvalue-decomposition-based exponentiation.
These are calculated via
\begin{equation}
    \hU = \exp(-i\hH) = \hV \exp(-i\hE) \hV^\dagger
\end{equation}
with eigenvalues $\mathbf{E}=\text{diag}(\hE)$ and eigenvector matrices $\hV$ determined for $\hH$ via the Lapack function \texttt{zheevd} \cite{lapack99} for the eigenvalue decomposition of hermitian matrices,
which utilizes the divide and conquer method optimized.

\section{Cache Size} \label{sec:cache_size}
We want to analyze the computational demands of creating the cache, and the memory requirements of the cache.
We can count the number of cached matrices from the grid.
For every edge $\bi\pm \bs_1$ along the first parameter direction we store one $\hat{L}_{\bi\pm\bs_1}$ matrix,
likewise we store one $\hat{R}_{\bi\pm\bs_n}$ matrix for every edge $\bi\pm \bs_n$ along the last parameter direction, leading to
\begin{equation}
    \# \hat{L} = N_1 \prod_{i=2}^{n} (N_i+1) \quad \mathrm{and} \quad \# \hat{R} = N_n \prod_{i=1}^{n-1} (N_i+1).
\end{equation}
For every grid point $\bi$ we store as many $\hat{C}_{\bi, \bs_p, \bs_{p+1}}$ matrices as there are quadrants next to it, so that the number
of cached matrices is given by the number of odd-summed grid points around a (hyper-) cubic unit cell $2^{n-1}$ times the number of quadrants $\prod_{i=1}^{n} N_i$
\begin{equation}
    \# \hat{C} = 2^{n-1} \prod_{i=1}^{n} N_i. \label{eq:number_central_cached_matrices}
\end{equation}
The overall number of cached matrices is then given by
\begin{align}
    C(\mathbf{N}) = & \# \hat{C} + \# \hat{L} + \# \hat{R} \nonumber                                             \\
    =               & 2^{n-1} \prod_{i=1}^{n} N_i + N_1 \prod_{i=2}^{n} (N_i+1) + N_n \prod_{i=1}^{n-1} (N_i+1).
\end{align}
In addition we store one eigenvalue vector $\bE_{\bi, \bs_p}$ for every edge $\bi \rightarrow \bi+\bs_p$ of the grid, leading to a cummulative $\# \bE = \prod_{i=1}^{n} N_i$.
The cache is constructed by
\begin{enumerate}
    \item Computing the grid point unitaries $\hU_\bi$ for all grid points $\bi$.
    \item For every edge $\bi+\bs_p$ of the grid, eigendecompose the product of the edges vertex unitaries as defined in Eq.~\eqref{eq:nd_ui_cached_matrices}
          to obtain the eigenenergies $\bE$ and eigenvectors $\hV$.
    \item Construct the cached matrices $\hat{L}$, $\hat{C}$ and $\hat{R}$ as defined in Eq.~\eqref{eq:control_param2interpolation_param_nd}.
\end{enumerate}
\section{Optimal Binning} \label{sec:optimal_caching}
In practice we wish to approximate our unitaries with guaranteed target precision $I_\mathrm{tar}$, while requiring the least amount of caching.
To facilitate this we approximate the maximum infidelity of different binnings $\mathbf{N}$ as a polynomial function of the bin widths
$\Delta c_i(N_i) = \frac{c_{i, \mathrm{max}} - c_{i, \mathrm{min}}}{N_i}$, using constants derived from a test binning
(for example a $N_i = 1 \, \forall i\in  \{1,\dots,n\}$). We then procede to develop an optimization routine.

We expand the average infidelity in Eq.~\eqref{eq:average_infidelity} of the UI into a Taylor expansion, using Eq.~\eqref{eq:nd_ui_bch}.
The leading order corrections in the exponent of the overlap operators $\hat{M}$ are of type
\begin{align}
    \hat{M} = & \hat{U}_\mathrm{exact}^\dagger(\mathbf{\alpha}) \hat{U}_\mathrm{UI}(\mathbf{\alpha}, \mathbf{N})  \nonumber \\
    =         & \mathrm{exp}(\sum_{i, j=1}^{n} \frac{1}{N_i N_j}  \alpha_i \alpha_j \hat{A}_{ij} + \mathcal{O}(N^{-3})),
\end{align}
where the $\hat{A}_{ij}$ are linear combinations of commutator terms involving a single $\hH_i$, $\hH_j$ and different numbers of $\hH_0$.
\footnote{The 1D case represents an exception, where $i=j=1$ and the commutators involved have two $\hH_i$  and at least one $\hH_0$ term.}
To estimate the maximum infidelity of the interpolation, we set $\alpha_i=\frac{1}{2} \, \forall i \in  \{1,\dots,n\}$.
The trace $\Tr(\hat{A}_{ij}) = 0$ must be zero, as all terms in $\hat{A}_{ij}$ are constructed from commutators, and commutators are traceless.
Therefore in leading order Taylor expansion the trace is given by
\begin{equation}
    \begin{aligned}
        \Tr(\hat{M}) = d + & \!\sum_{\substack{i<j \\k<l}}^{n}\! \frac{1}{16 N_i N_j N_k N_l} \Tr(\hat{A}_{ij} \hat{A}_{kl}) + \\
        +                  & \mathcal{O}(N^{-6}).
    \end{aligned}
\end{equation}

In Appendix \ref{sec:traceless_traces} we show that the traces $\Tr(\hat{A}_{ij} \hat{A}_{kl})$ are dominated by terms with $i,j=k,l$, while terms $i,j \neq k,l$ are supressed.
We therefore neglect these terms in our expansion of the trace of $\hat{M}$.
\begin{equation}
    \Tr(\hat{M}) = d + \sum_{1\leq i<j}^{n} \frac{1}{16 N_i^2 N_j^2} \Tr(\hat{A}_{ij}^2) + \mathcal{O}(\Delta c^6).
\end{equation}
Lacking an efficient way to compute the terms $\hat{A}_{ij}$ or even just generally deriving the commutator terms, we instead approximate at the
point of maximum infidelity using
\begin{equation}
    \begin{aligned}
        \hat{A}_{ij} \approx & \hat{U}^\dagger_\mathrm{exact}(\alpha_i=\alpha_j=\frac{1}{2}, \alpha_{k\neq i,j}=0)\cdot                              \\
        \cdot                & \hat{U}_\mathrm{UI}(\alpha_i=\alpha_j=\frac{1}{2}, \alpha_{k\neq i,j}=0, N_i = 1))-                                   \\
        -                    & \mathds{I} - \begin{cases} 0 & i=j \\ \hat{A}_{ii}-\hat{A}_{jj} & i\neq j \end{cases}. \label{eq:approx_unitary_to_A}
    \end{aligned}
\end{equation}
Here we remove other terms depending only on $i$ or $j$ from the approximation,
as they are also included in the unitaries but depend on $\frac{1}{N_i^4}$.
We cache the traces $T_{ij}= 2\mathrm{Re}(\Tr(\hat{A}_{ij}^2))$ once at the beginning of the procedure,
to then repeatedly approximate the infidelity as
\begin{equation}
    I(\mathbf{N}) \approx \frac{1}{d+1}\sum_{i \leq j}^{n} \frac{1}{N_i^2 N_j^2}   T_{ij} + \mathcal{O}\Big( \frac{1}{N^6} \Big), \label{eq:infidelity_bin_approximation}
\end{equation}
to optimize the number of bins $\mathbf{N}$.

\subsection{Optimization of Binning}
The bin optimization is divided into three steps, first we increase the number of bins $\mathbf{N}$ to achieve an estimated infidelity
below the designated threshold $I(\mathbf{N}) < I_\mathrm{tar}$,
we then decrease the number of bins in two steps, to minimize the number of bins $C(\mathbf{N})$ while keeping the infidelity below the target infidelity. \\

\subsubsection{Increasing Cache Size to Reach a Target Infidelity}
We start with an initial binning $N_i^{(0)} = 1 \, \forall i \in  \{1,\dots,n\}$. In every iteration of the optimization, we increase the bin count by one
$N_i^{(k+1)} = N_i^{(k)}+1$ for one-parameter direction $i$, and keep all other directions constant $N_h^{(k+1)}=N_h^{(k)} \forall h\neq i$.
To this end, for every direction $j \in  \{1,\dots,n\}$, we calculate the ratio $\mathcal{L}_j$ of the infidelity decrease to the cache size increase
\begin{equation}
    \mathcal{L}_j(\mathbf{N}^{(k)})  = \frac{I(\mathbf{N}^{(k)}) - I(\mathbf{N}^{(k)}+\mathbf{e}_j)}{C(\mathbf{N}^{(k)} + \mathbf{e}_j) - C(\mathbf{N}^{(k)})},
\end{equation}
and choose the direction $i$ with the largest ratio
\begin{equation}
    i= \mathrm{argmax}_j \mathcal{L}_j(\mathbf{N}),
\end{equation}
for the increase of $N_i+1$. 
The procedure ends, once the estimated infidelity is below the target infidelity $I(\mathbf{N}^{(k)}) < I_\mathrm{tar}$.

\subsubsection{Decreasing Cache Size while Staying Below a Target Infidelity}
In every optimization step we decrease the bin number by one in one direction $\mathbf{N}_i^{(k+1)} = \mathbf{N}_i^{(k)} - 1$,
where we choose the direction $i$ with the largest decrease in cache size, which still keeps the infidelity below the target infidelity
\begin{equation}
    i = \mathrm{argmin}_i C(\mathbf{N} - \mathbf{e}_i) \quad  \forall i: I(\mathbf{N} - \mathbf{e}_i) < I_\mathrm{tar}.
\end{equation}
This procedure ends, when the current binning $\mathbf{N}$ is the smallest binning, which still keeps the infidelity below the target infidelity $I(\mathbf{N}) < I_\mathrm{tar}$.
In our tests, this procedure consistently found optima that were within $1\%$ of the optimal binning, as determined by exhaustive search, while reducing the number of bins by $50\%$.

\subsubsection{Further Decrease in Cache Size}
We further refine the previous step, by utilizing a constrained analytical solution to find the number of bins in direction $m$,
so that the estimated infidelity $I(\mathbf{N})$ is equal to the target infidelity $I_\mathrm{tar}$, keeping the other binning $N_{j \neq m}$ constant.
Using Eq.~\eqref{eq:infidelity_bin_approximation} and substituting both $s_i = \frac{1}{N_i^2}$ and $\frac{T_{ij}}{d(d+1))}= P_{ij}$, we can rewrite the target infidelity condition as
\begin{align}
    0 = & I(\mathbf{s}) - I_\mathrm{tar} = \sum_{i<j}^{n} s_i s_j P_{ij} + \sum_{i=1}^{n} s_i^2 P_{ii}  - I_\mathrm{tar} \nonumber \\
    =   & \underbrace{\sum_{\substack{i<j                                                                                          \\ i \neq m \neq j}}^{n} s_i s_j P_{ij} + \sum_{i \neq m}^{n} s_i^2 P_{ii} - I_\mathrm{tar}}_{=I_\mathrm{min}-I_\mathrm{tar}=I_0}
    + s_m\underbrace{\sum_{j\neq m}^{n} s_j P_{mj}}_{=Q_m} + s_m^2 P_{mm}. \label{eq:infidelity_bin_approximation_s}
\end{align}
We solve this quadratic equation, so that
\begin{equation}
    s_{m, \pm} = - \frac{Q_m}{2 P_{mm}} \pm \sqrt{\left(\frac{Q_m}{2 P_{mm}}\right)^2 - \frac{I_0}{P_{mm}}}. \label{eq:bin_optimization_s}
\end{equation}
This equation only has relevant solutions, if $I_\mathrm{min} < I_\mathrm{tar} \rightarrow I_0 < 0$.
We are furthermore interested in the smallest positive solution of $s_{m, \pm}$, which we will call $s_{m, \mathrm{min}}$,
and if such a solution exists, determine $N_m = \frac{1}{s_{m, \mathrm{min}}^2}$.

We can now optimize by test-wise reduction of one of the bin numbers by one $N_i^{(k+1)}=N_i^{(k)}-1$ and then calculating the optimal bin number $N_m^(N_{i\neq m}^{(k+1)})$ for each of the other directions $m$.
We repeat this for every test-wise reduction of one bin number, and choose the reduction combination $i,m$, which leads to the largest decrease in cache size,
while keeping the infidelity below the target infidelity. We repeat this until no reductions can be found.

\subsection{Traces of Products of Traceless Operators}\label{sec:traceless_traces}
We consider the traceless operators $A_{ij}$ and $A_{kl}$ from section \ref{sec:optimal_caching},
and wish to show, that terms with $ij \neq kl$ are statistically significantly smaller than terms with $ij=kl$.
To show this, let us expand the operators into the generalized Pauli matrices ($n$-qubit Pauli matrices)\footnote{For systems that cannot be decomposed into qubits, we can simply add dimensions with zero valued entries in $\hat{A}_{ij}$ and $\hat{A}_{kl}$.}
\begin{equation}
    \hat{A}_{ij} = \sum_{\mu=0}^{4^{n}-1} a_\mu \hat{\sigma}_\mu, \quad \mathrm{and} \quad \hat{A}_{kl} = \sum_{\nu=0}^{4^{n}-1} b_\nu \hat{\sigma}_\nu,
\end{equation}
here the particular indexing of the Pauli matrices is not important, as long as the same indexing is used for both operators.
We find that the trace of the product of the two operators is given by
\begin{equation}
    \Tr(\hat{A}_{ij} \hat{A}_{kl}) = \sum_{\mu, \nu=0}^{4^{n}-1} a_\mu b_\nu \Tr(\hat{\sigma}_\mu \hat{\sigma}_\nu) = \sum_{\mu=0}^{4^{n}-1} a_\mu b_\mu \Tr(\hat{\sigma}_\mu^2),
\end{equation}
where we used the fact, that the trace of the product of two Pauli matrices is zero $\Tr(\hat{\sigma}_\mu \hat{\sigma}_\nu) = d \delta_{\mu, \nu}$.
We therefore conclude, that
\begin{equation}
    \Tr(\hat{A}_{ij} \hat{A}_{kl}) = 2 d\|a\| \|b\| \cos(\theta)
\end{equation}
For diagonal terms $ij=kl$ we have $\cos(\theta) = 1$, so that $\Tr(A_{ij}A_{ij})=d\|a\|^2$, while for off-diagonal terms $ij \neq kl$ the result
will depend on the angle between the matrices. We assume, that the Pauli vectors $\mathbf{a}$ and $\mathbf{b}$ are randomly distributed,
in table \ref{tab:traceless_traces} we show this to also be a valid approximation for trace-orthogonal Hamiltonian terms $\Tr(\hH_i \hH_j)=0 \, \forall i \neq j$.
The cosine of the angle $\cos(\theta)=$ between two random vectors $\mathbf{a}$ and $\mathbf{b}$ isotropically sampled from the unit (hyper) sphere,
using the n-dimensional isotropic distribution \cite{Cai2013}
\begin{equation}
    p(\cos \theta) = \frac{\Gamma(\frac{n}{2})}{\sqrt{\pi} \Gamma(\frac{n-1}{2})} (1 - \cos^2 \theta)^{\frac{n-3}{2}} \quad \text{for } -1 \leq \cos \theta \leq 1,
\end{equation}
with expectation value $\braket{\cos(\theta)}=0$ and variance $\mathrm{Var}(\cos(\theta)) = \braket{\cos(\theta)^2}=\frac{1}{d^2}$.
We therefore conclude, that the trace of the off-diagonal terms will be
$\mathds{E}(\Tr(A_{ij}A_{kl}))=0$ with variance $\mathrm{Var}(\Tr(A_{ij}A_{kl}))=\frac{1}{d^2}\|a\|\|b\|$.

\subsection{BCH Expansion of Unitary Interpolations} \label{sec:ui_bch}
\subsubsection{One-Parametric Unitary Interpolation} \label{sec:ui_1d_bch}
\begin{equation}
    \begin{aligned}
        \hU_\UI & (\alpha) = \exp\Big(-i\Big(\overbrace{\hH_{0} + \Big(c_0 + \alpha \frac{\Delta c}{N}\Big) \hH_{1}}^{\text{Desired Terms}} -                                                                                                \\
        -       & \underbrace{\frac{\Delta c^2}{12 N^2} \left( \alpha^2 -\alpha\right) \left[\left[\hH_{0},\hH_{1}\right],\hH_{1}\right] + \bigO\Big(\frac{\Delta c^2}{N^2}\Big)}_{\text{Undesired Terms}} \Big) \Big). \label{eq:ui_1d_bch}
    \end{aligned}
\end{equation}
\subsubsection{Multi Parametric Unitary Interpolation} \label{sec:ui_nd_bch}
\begin{equation}
    \begin{aligned}
        \hU_\UI (\balpha) & = \exp\Big( -i\Big(\overbrace{\hH_0+ \sum_{m=1}^{n} \frac{c_k}{N_k} \hH_m}^{\mathrm{Exact Hamiltonian}}\Big) +                                                                                                  \\
        +                 & \underbrace{\frac{1}{2}\sum_{ 1 \leq p < q}^{n} \alpha_p \alpha_q \frac{\Delta c_p \Delta c_q}{N_p N_q} \comb{p}{q} + \bigO\Big(\frac{\Delta c^2}{N^2}\Big)}_{\mathrm{Corrections}} \Big), \label{eq:nd_ui_bch}
    \end{aligned}
\end{equation}

\section{BCH Expansions of Trotterisations} 
\subsection{Suzuki-Trotter} \label{sec:trotter_bch}
The BCH expansion of a time step $\Delta t$ with into $N$ Trotter steps, of an $n$-parameter problem, reveals
\begin{equation}
    \begin{aligned}
         & \hU_\Tr (\bc) = \Big(\Big(\prod_{p=1}^{n}\exp\Big(-i\frac{\hH_p}{N} c_p\Big)\Big) \exp\Big(-i\frac{\hH_0}{N} \Big)\Big)^N \\
         & \,=\exp\Big( -i\Big(\hH_{0} + \sum_{p=1}^{n} c_p \hH_p\Big) -\frac{1}{2N} \Big(\sum_{p=1}^{n} c_p \comb{0}{p} +           \\
         & \,+ \sum_{1\leq p < q}^{n} c_p c_q \comb{p}{q}\Big) + \mathcal{O}\Big(\frac{c}{N}\Big)\Big)\Big). \label{eq:trotter_bch}
    \end{aligned}
\end{equation}
Each of the parametrized exponentials can be eigendecomposed (as in the UI) $\bE_i,\, \hV_i = \eig(\hH_i)$.
The product of the adjacent eigenvector matrices in the construction of the expression in Eq.~\eqref{eq:trotter_bch} can be precomputed
$\hat{C}_i = \hV_{i+1}^\dagger \hV_i$ and cached together with the eigenvalues $\bE_i$.
A single Trotter step then requires the same number of matrix multiplications as the UI.
$N=2^j$ Trotter steps can be computed from a single Trotter step $\hU_\Tr$ by repeated squaring $\hU_\Tr^{2^j} = \hU_\Tr^{2^{j-1}} \hU_\Tr^{2^{j-1}}$.

\subsection{Symmetric Trotter (Strang or Split-Step))} \label{sec:strang_bch}
The symmetrized split-step variant of the Trotter method is given by
\begin{equation}
    \begin{aligned}
        \hU_\St (c) = & \Big(\Big(\prod_{p=1}^{n}\exp\Big(-i\frac{\hH_p}{2N} c_p\Big)\Big)\exp(-i\frac{\hH_0}{N})
        \Big(\coprod_{p=1}^{n}\exp(-i\frac{hH_p}{2N} c_p)\Big)\Big)^N = \exp\Big( -i\Big(\hH_{0} + \sum_{p=1}^{n} c_p \hH_p + \\
        +             & \sum_{p=1}^{n} \frac{c_p}{12N^2}\left(\comt{0}{p}{0}+ \frac{c_p}{2} \comt{0}{p}{p} \right)
        + \sum_{1\leq p < q}^{n} \frac{c_p c_q}{12N^2}\left( \frac{c_q}{2}\comt{p}{q}{q} + c_p \comt{p}{q}{p}\right) +        \\
        +             & \sum_{0 \leq p < q < r}^{n} \frac{c_p c_q c_r}{12N^2}\left(\comt{p}{r}{q} + \comt{q}{r}{p} \right) +
            \bigO\Big(\frac{c}{N^2}\Big)\Big)\Big), \label{eq:strang_bch}
    \end{aligned}
\end{equation}
where the coproduct sign $\coprod_{p=1}^{n}$ denotes the reversed order of the product.
This adds $n-1$ additional matrix multiplication over Trotterisation and UI for the construction of a single symmetric Trotter step, while the repeated squaring remains the same.

\section{Symmetric Unitary Interpolation}\label{sec:sym_ui_bch}
In the main part we also compared the UI with a symmetrized version of the UI.
As this methods added accuracy did not justify the additional computational cost,
we did not include it in the main text. Here we quickly introduce the method.
Using the indexing and interpolation parameters from Section \ref{sec:nd_interpolation},
we can generalize the UI akin to the split step approach by symmetrization.
This requires half step unitaries on the interpolation grid
\begin{equation}
    \tU_\ft{\mathbf{i}} = \exp\left(-\frac{i}{2}\left(\hH_0+\sum_{p=1}^{n} \Big(c_{p,\mathrm{min}} + i_p \frac{\Delta c_p}{N_p}\Big) \hH_p\right)\right). \label{eq:half_steps_for_h_elements}
\end{equation}

The interpolation is then constructed as
\begin{equation}
    \begin{gathered}
        \hU_{\UI,S}(\bc, \balpha(\bc)) = \tU_\ft{\bi} \left(\prod_{p=2}^{n} \left( \tU_\ft{\bi}^\dagger \tU_\ft{\bi+\bs_p} \right)^{|\alpha_p|}\right)
        \left( \tU_\ft{\bi}^\dagger \tU_\ft{\bi+\bs_1} \tU_\ft{\bi}^\dagger \right)^{|\alpha_1|}
        \left(\coprod_{p=2}^{n} \left(  \tU_\ft{\bi+\bs_p} \tU_\ft{\bi}^\dagger \right)^{|\alpha_p|}\right) \tU_\ft{\bi}. \label{eq:sym_ui}
    \end{gathered}
\end{equation}
Here the $\coprod$ symbol denotes the product of the matrices in the reverse order (from the right side).
The lattice indexes $\bi$ and interpolation parameters $\balpha$ are calculated analogously to the case of the UI in Eq.~\ref{eq:nd_ui}.
This product again can be constructed from cached matrices which are derived from the logarithmic decomposition
of the products $\left(  \tU_\ft{\bi+\bs_p} \tU_\ft{\bi}^\dagger \right)=\hat{L}_{\bi,\bs_p} \exp(-i\bE_{\bi+\bs_p}) \hat{L}_{\bi,\bs_p}^\dagger$,
and their counterparts $\left(  \tU_\ft{\bi+\bs_p} \tU_\ft{\bi}^\dagger \right)=\hat{R}_{\bi,\bs_p}^\dagger \exp(-i\bE_{\bi+\bs_p}) \hat{R}_{\bi,\bs_p}$,
giving us $\hat{C}^{(L)}_{\bi, \bs_{p+1},\bs_p} = \hat{L}_{\bi,\bs_{p+1}}^\dagger\hat{L}_{\bi,\bs_p}$ and
$\hat{C}^{(R)}_{\bi, \bs_{p+1},\bs_p}= \hat{R}_{\bi,\bs_p}\hat{R}_{\bi,\bs_{p+1}}^\dagger$ respectively,
doubling the number of central cached matrices in Eq.~\eqref{eq:number_central_cached_matrices}.\\
\subsection{BCH expansion}
By repeated BCH expansion we find this to result in (where we used $d_p = c_{p,\mathrm{min}}+i_p\frac{\Delta c_p}{N_p}$)
\begin{equation}
    \begin{aligned}
        \hU_{\UI,S}(\bc, \balpha(\bc)) & = \exp\Big(-i\Big( \hH_{0} - \sum_{p=1}^{n} c_p \hH_p - \frac{1}{48}\sum_{p=1}^{n} \frac{\Delta c_p^2}{N_p^2} (|\alpha_p|^2-|\alpha_p|)\comt{0}{p}{p} +                                                                                                       \\
                                       & - \frac{1}{24}\sum_{1 \leq p < q}^{n} |\alpha_p| |\alpha_q| \frac{\Delta c_p\Delta c_q}{N_p N_q} \left( 2\comt{0}{p}{q} - \comt{0}{q}{p}\right) +                                                                                                             \\
                                       & - \frac{1}{48}\sum_{1 \leq p < q}^{n} \left( d_q \Delta \frac{c_p^2}{N_p^2}(|\alpha_p| - |\alpha_p|^2) - 2|\alpha_q| \frac{\Delta c_q}{N_q}\Big(d_p |\alpha_p|\frac{\Delta c_p}{N_p} + 2 \frac{(|\alpha_p| \Delta c_p)^2}{N_p^2}\Big) \right)\comt{p}{q}{p} - \\
                                       & + \frac{1}{48}\sum_{1 \leq p < q}^{n} \left( d_p \Delta \frac{c_q^2}{N_q^2}(|\alpha_q| - |\alpha_q|^2) + 2|\alpha_p| \frac{\Delta c_p}{N_p}\Big(2 d_q |\alpha_q|\frac{\Delta c_q}{N_q} + \frac{(|\alpha_q| \Delta c_q)^2}{N_q^2}\Big) \right)\comt{p}{q}{q} + \\
                                       & - \frac{1}{12}\sum_{1 \leq p < q < r}^{n} |\alpha_r|\frac{\Delta c_r}{N_r}\left( |\alpha_p|\frac{\Delta c_p}{N_p} d_q  - d_p |\alpha_q|\frac{\Delta c_q}{N_q}  -  \alpha_p\frac{\Delta c_p}{N_p} |\alpha_q| \frac{\Delta c_q}{N_q} \right)\comt{p}{q}{r} -    \\
                                       & + \frac{1}{24}\sum_{1 \leq p < q < r}^{n} |\alpha_q|\frac{\Delta c_q}{N_q}\left( d_p |\alpha_r| \frac{\Delta c_r}{N_r} +2 |\alpha_p| \frac{\Delta c_p}{N_p} d_r\right)\comt{p}{r}{q} +                                                                        \\
                                       & - \frac{1}{24}\sum_{1 \leq p < q < r}^{n} |\alpha_p|\frac{\Delta c_p}{N_p}\left( d_q |\alpha_r| \frac{\Delta c_r}{N_r} - |\alpha_q| \frac{\Delta c_q}{N_q} d_r\right)\comt{q}{r}{p} + \bigO\Big(\frac{\Delta c^2}{N^2}\Big)\Big)\Big) \label{eq:sym_ui_bch}
    \end{aligned}
\end{equation}
Like in the symmetric Trotterisation, this results in a method with leading order infidelity $I \propto \bigO\Big( \frac{1}{N^2}\Big)$, only outperforming the UI by a constant ratio, due to better coefficients in the expansion.

\end{document}